\documentclass[11pt,a4paper]{article}
\pdfoutput=1
\usepackage[usenames, dvipsnames]{color}
\usepackage[compat=1.1.0]{tikz-feynman}
\usepackage{jheppub}
\usepackage{amssymb,amsmath}
\usepackage{url,booktabs}
\usepackage{slashed}
\usepackage{multirow}
\usepackage[english=american]{csquotes}
\usepackage{color,soul}
\usepackage{caption}
\usepackage{mathtools}
\usepackage{braket}
\usepackage{relsize} 
\usepackage{subcaption}
\usepackage{xfrac}
\usepackage{enumitem} 

\DeclareMathOperator{\keV}{keV}

\DeclareMathOperator{\GeV}{GeV}
\DeclareMathOperator{\TeV}{TeV}

\DeclareMathOperator{\LH}{L}

\DeclareMathOperator{\PRH}{P_R}
\DeclareMathOperator{\PLH}{P_L}

\def\be{\begin{equation}}
\def\ee{\end{equation}}
\def\bea{\begin{eqnarray}}
\def\eea{\end{eqnarray}}

\graphicspath{{Plots/}}

\title{Mechanism for baryogenesis via feebly interacting massive particles}

\author[1]{Andreas Goudelis}
\author[2]{\!\!, Pantelis Papachristou}
\author[2]{\!\!, Vassilis C. Spanos}
\affiliation[1]{Laboratoire de Physique de Clermont (UMR 6533), CNRS/IN2P3, Univ.\ Clermont Auvergne, 4 Av.\ Blaise Pascal, F-63178 Aubi\`ere Cedex, France}
\affiliation[2]{National and Kapodistrian University of Athens, Department of Physics, Section of Nuclear \& Particle Physics, GR-157 84 Athens, Greece}
\emailAdd{andreas.goudelis@clermont.in2p3.fr}
\emailAdd{pantelisp@phys.uoa.gr}
\emailAdd{vspanos@phys.uoa.gr}

\abstract{%
We present a simple mechanism which allows the simultaneous generation of the baryon asymmetry of the Universe along with its dark matter content. To this goal, we employ the out-of-equilibrium decays of heavy bath states into a feebly coupled dark matter particle and Standard Model charged fermions. These decays lead to dark matter production via the freeze-in mechanism and, assuming that they further violate $CP$, can generate a viable matter-antimatter asymmetry in the resonant regime. We illustrate this mechanism by studying  a particular realization of this general scenario, where the role of the heavy bath particles is played by $SU(3)_{\text{c}}\times SU(2)_{\LH}$-singlet vectorlike fermions with a non-zero hypercharge and dark matter is identified with a gauge-singlet real scalar field. We show that in the context of this simple model the cosmological constraints for the dark matter abundance and the baryon asymmetry are satisfied for masses of heavy vectorlike fermion states of a few TeV, potentially within reach of the High-Luminosity Run of the Large Hadron Collider. Dark matter, in turn, is predicted to be rather light, with a mass of a few keV.
}

\begin{document}

\maketitle
\section{Introduction}
\label{sec:intro}

Among the numerous open questions in contemporary high-energy physics, the origin of cosmic dark matter (DM) and that of the baryon asymmetry in the Universe occupy a pivotal place. Not only do they constitute two of the most striking and fundamental pieces of evidence for the existence of physics beyond the Standard Model (BSM) of particle physics, but they do so only once the latter is placed within a cosmological framework. Resolving either (let alone both) of these questions most likely requires particle physics to be viewed from a cosmological standpoint and, conversely, cosmology to be analyzed in terms of the behavior of the fundamental constituents of matter and their interactions. And, indeed, in doing so during the past decades model builders have not been short of ideas concerning the nature of dark matter and the mechanism through which matter came to dominate over antimatter in the Universe. 

Among the various dark matter candidates that have been proposed we can mention axions \cite{Preskill:1982cy, Abbott:1982af, Dine:1982ah}, primordial black holes \cite{Hawking:1971ei, Chapline:1975ojl, Green:2020jor}, a vast number of incarnations of weakly or feebly interacting massive particles (WIMPs or FIMPs; for reviews \textit{cf e.g.} \cite{Arcadi:2017kky, Bernal:2017kxu}), gravitationally produced dark matter \cite{Kolb:1998ki}, and asymmetric dark matter (for reviews \textit{cf e.g.} \cite{Davoudiasl_2012, Petraki:2013wwa}), just to name a few. Similarly, the matter-antimatter asymmetry of the Universe has been explained in terms of different mechanisms of baryogenesis, like baryogenesis based on grand unified theories (GUTs) \cite{Yoshimura:1978ex}, electroweak baryogenesis \cite{Kuzmin:1985mm,Cohen:1993nk}, the Affleck-Dine mechanism \cite{Affleck:1984fy}), and leptogenesis \cite{Fukugita:1986hr}. Interestingly, during the past decade, there have also been several attempts to actually link the two questions, most notably in the contexts of WIMP baryogenesis \cite{McDonald_2011,Cui_2012,Cui_2013} or asymmetric dark matter \cite{Hall:2010jx,Unwin:2014poa,Cui:2020dly}.

Recently, the main idea behind Akhmedov-Rubakov-Smirnov (ARS) leptogenesis \cite{Akhmedov:1998qx}, namely, that of a lepton number asymmetry being generated through $CP$-violating sterile neutrino oscillations, was exploited in \cite{Shuve_2020}, where the authors proposed that a baryon asymmetry could, instead, be also generated by augmenting the SM with exotic scalars and fermions directly coupling to quarks. The fermions, which were taken to be singlets under the SM gauge group can, moreover, play the role of viable dark matter candidates through the freeze-in mechanism \cite{McDonald:2001vt, Hall:2009bx}, whereas their $CP$-violating oscillations, in the presence of electroweak sphaleron transitions, can generate the observed baryon asymmetry of the Universe.

In this paper, we propose a mechanism which borrows ideas both from Dirac leptogenesis \cite{Dick:1999je,Murayama_2002,Cerdeno_2006,Gonzalez:2009} and from this scenario of ``freeze-in baryogenesis''. As we will describe in detail in the following, we consider a heavy particle species which is charged under (parts of) the SM gauge group and which can decay through feeble interactions into SM fermions along with a neutral particle. The latter is our dark matter candidate, produced upon the decays of the heavy particle through the freeze-in mechanism, along the lines presented in \cite{Shuve_2020}. However, in our case the decays \textit{themselves} violate $CP$, in a similar manner as in leptogenesis models. Then, in the presence of electroweak sphalerons, we will see that an asymmetry can be generated between SM fermions and antifermions. The relevant processes proceed through feeble couplings, preventing them from ever reaching equilibrium and, thus, satisfying the third Sakharov condition \cite{Sakharov:1967dj}. The first condition, namely, baryon number violation, is satisfied due to the active sphaleron processes in a way that resembles neither GUT baryogenesis (explicit $B$ violation in decays) nor leptogenesis (explicit $L$ violation in decays), although we will also comment on the possibility of direct $B/L$ violation as well.

The paper is structured as follows: in Section \ref{sec:GeneralFramework} we discuss the general features of our mechanism, namely, the way through which the required dark matter abundance and the baryon/lepton asymmetries can be generated, without adopting any concrete microscopic model. In Section \ref{sec:Model} we propose a simple model as a proof-of-concept that concrete incarnations of our mechanism can, indeed, be constructed. We compute the predicted dark matter abundance and baryon asymmetry, quantify the effects of processes that wash out the latter and briefly comment on the phenomenological perspectives of our model, notably in relation to searches for long-lived particles (LLPs) at the Large Hadron Collider (LHC). Finally, in Section \ref{sec:conclusions} we summarize our main findings and conclude. Some more technical aspects are left for the Appendix.

\section{Dark matter and baryogenesis from freeze-in: General framework}\label{sec:GeneralFramework}

Before presenting a concrete realization of our take on the freeze-in baryogenesis idea, let us briefly summarize a few key notions that will be useful for the discussion that follows: frozen-in dark matter and how the freeze-in framework can enable us to satisfy the three Sakharov conditions that are necessary for successful baryogenesis. A concrete implementation of these ideas will be presented in detail in Section \ref{sec:Model}.

\subsection{Freeze-in DM abundance}\label{Freeze-in DM abundance}

The freeze-in mechanism for dark matter production relies on two basic premises: 

\begin{itemize}
\item The initial DM abundance is zero.
\item Dark matter interacts only extremely weakly (``feebly'') with the Standard Model particles (along with any other particles that are in thermal equilibrium with them).
\end{itemize}
Under these assumptions, and further assuming that dark matter production takes place in a radiation-dominated Universe, dark matter never reaches thermal equilibrium with the plasma. Instead, it is produced through the out-of-equilibrium decays or annihilations of bath particles and all dark matter depletion processes, the rate of which typically scales as $\left\langle \sigma v \right\rangle \times n_{DM}^2$ (where $\sigma$ is the dark matter annihilation cross section, $v$ its velocity, and $n_{DM}$ its number density), can be ignored.

\subsection{Baryon asymmetry abundance $Y_{B}$}\label{sec:Basymmetrygeneral}

In general, the decays and/or annihilations that are responsible for dark matter production can also violate both the baryon number $B$ and $C/CP$.\footnote{Very similar remarks can be made if the decay violates, instead, the lepton number. This is also the option that we will adopt later in this work.} Then, as long as we stick to the freeze-in framework, these processes occur out-of-equilibrium with the thermal plasma, thus fulfilling all three Sakharov conditions.

Intuitively, and ignoring all washout processes, if we denote the measure of $CP$ violation by $\epsilon_{CP}$, we would expect the generated asymmetry in the SM fermion $Y_{\Delta f}$ to be connected to the dark matter abundance $Y_{DM}$ through a relation of the type
\begin{equation}\label{eq:BasymmetryDM}
Y_{\Delta f}\,\sim\,\epsilon_{CP}\,Y_{DM}
\end{equation}
In reality, this limit cannot be attained given that some amount of washout is almost inevitable, whereas concrete realizations typically require the introduction of additional particles and decay channels. In this respect this relation may be viewed as an upper limit to the asymmetry that can be generated through decays that simultaneously produce dark matter.

In fact, in the following, when studying a concrete incarnation of our decay-induced freeze-in baryogenesis idea, we will see the following.
\begin{itemize}
    \item Since we will be starting with non-self-conjugate initial states $F_i$ (Dirac fermions), $CPT$ conservation and unitarity impose the existence of multiple decay channels of the $F_i$'s for a non-vanishing $CP$ asymmetry to be generated. To this goal, we will exploit possible decays of the heavy fermions into different Standard Model fermions (leptons), \textit{i.e.} flavor effects.
    
    \item The freeze-in framework will necessitate extremely small values for the couplings involved in the decay process. The predicted $CP$ violation, being an effect that arises from the interference of tree-level and one-loop processes is, then, even further suppressed, which will lead us to consider resonantly enhanced configurations in self-energy-type diagrams \cite{Pilaftsis_2004, Pilaftsis:2005rv, Pilaftsis_1999, Anisimov_2006}. Therefore, at least two heavy fermions $F_i$ must be added.
    
    \item The baryon and/or lepton number need not be violated by the decay processes. As proposed, \textit{e.g.} in \cite{Dick:1999je}, a $CP$ asymmetry can be translated to a baryon asymmetry by the electroweak sphalerons. If, additionally, the BSM heavy fermions are immune to the action of sphalerons, in the end a net asymmetry will be generated in the baryon and lepton sectors.
\end{itemize}

\section{A concrete realization}\label{sec:Model}

Let us now elaborate the previous considerations through a concrete, simple model. We extend the SM by two heavy vectorlike leptons $F_i$ which are singlets under $SU(3)_{\text{c}}\times SU(2)_{\LH}$ but carry hypercharge and a real gauge-singlet scalar $S$ which is our freeze-in DM candidate. We moreover impose a discrete $Z_2$ symmetry on the Lagrangian, under which all exotic states are taken to be odd while the SM particles are even. Under these assumptions, the Lagrangian reads

\begin{equation}\label{eq:Lgeneral}
\mathcal{L}=\mathcal{L}_{\text{SM}}+\mathcal{L}_{S}+\mathcal{L}_{SF}
\end{equation}

\noindent
where $\mathcal{L}_{\text{SM}}$ is the SM Lagrangian,

\begin{equation}\label{eq:LS}
\mathcal{L}_{S} = \partial_\mu S ~ \partial^\mu S  - \frac{\mu_S^2}{2} S^2 + \frac{\lambda_S}{4!} S^4 + \lambda_{Sh} S^2 \left(H^\dagger H\right)
\end{equation}

\noindent
describes interactions of dark matter with itself and with the Standard Model Higgs doublet and

\begin{align}
\mathcal{L}_{SF}\, =\, \sum_{i} \left( \bar{F}_i \left(i\slashed{D}\right) F_i  - M_i \bar{F_i} F_i \right) 
& -\, \sum_{\alpha, i} \left( \lambda_{\alpha i}\,S\,\bar{F}_i\,\PRH\,e_{\alpha}\,+\,\lambda_{\alpha i}^*\,S\,\bar{e}_{\alpha}\,\PLH\,F_i \right)
\end{align}

\noindent
where $e_{\alpha}$ are the right-handed SM charged leptons of flavor $\alpha=\{e,\mu,\tau\}$ and $\lambda_{\alpha i}$ denote the feeble couplings. The heavy fermions $F_i$ are assumed to carry the same lepton number with the SM leptons, so their interactions are lepton number conserving. Note that, without loss of generality, we have neglected potential off-diagonal couplings among the heavy fermions $F_i$. For simplicity in what follows, we will also set the Higgs portal coupling, $\lambda_{Sh}$, to zero.
\\
\\
The tree-level proper decay width $\Gamma_{F_i\rightarrow e_{\alpha}\,S}\big|_0$ in the limit $M_i\gg m_{e_{\alpha}}+m_{S}$ is given by

\begin{equation}\label{eq:decay width}
\Gamma_{F_i\rightarrow e_{\alpha}\,S}\big|_0\,\simeq\,\frac{\left|\lambda_{\alpha i}\right|^2}{16\pi\,g_{F_i}}\,M_i
\end{equation}

\noindent
where $g_{F_i}=2$ are the internal degrees of freedom of species $F_i$. The equilibrium decay rate density $\gamma_{F_i\rightarrow e_{\alpha}\,S}$ evaluated in the decaying particle rest frame is \cite{Hall:2009bx}

\begin{align}\label{eq:decay rate density}
\gamma_{F_i\rightarrow e_{\alpha}\,S}\,&\equiv\,\int\text{d}\Pi_{F_i}\,\text{d}\Pi_{e_{\alpha}}\,\text{d}\Pi_{S}\,\left(2\pi\right)^4\,\delta^{(4)}\left(p_{F_i}-p_{e_{\alpha}}-p_{S}\right)\,f^{\text{eq}}_{F_i}\,\left|\mathcal{M}\right|^2_{F_i\rightarrow e_{\alpha}\,S}\nonumber
\\
&=\,\frac{g_{F_i}}{2\pi^2}\,M_i^2\,\Gamma_{F_i\rightarrow e_{\alpha}\,S}\,T\,K_1\left(\frac{M_i}{T}\right)
\end{align}

\begin{figure}[t]
    \centering
\begin{subfigure}{0.24\textwidth}
  \includegraphics[width=\linewidth]{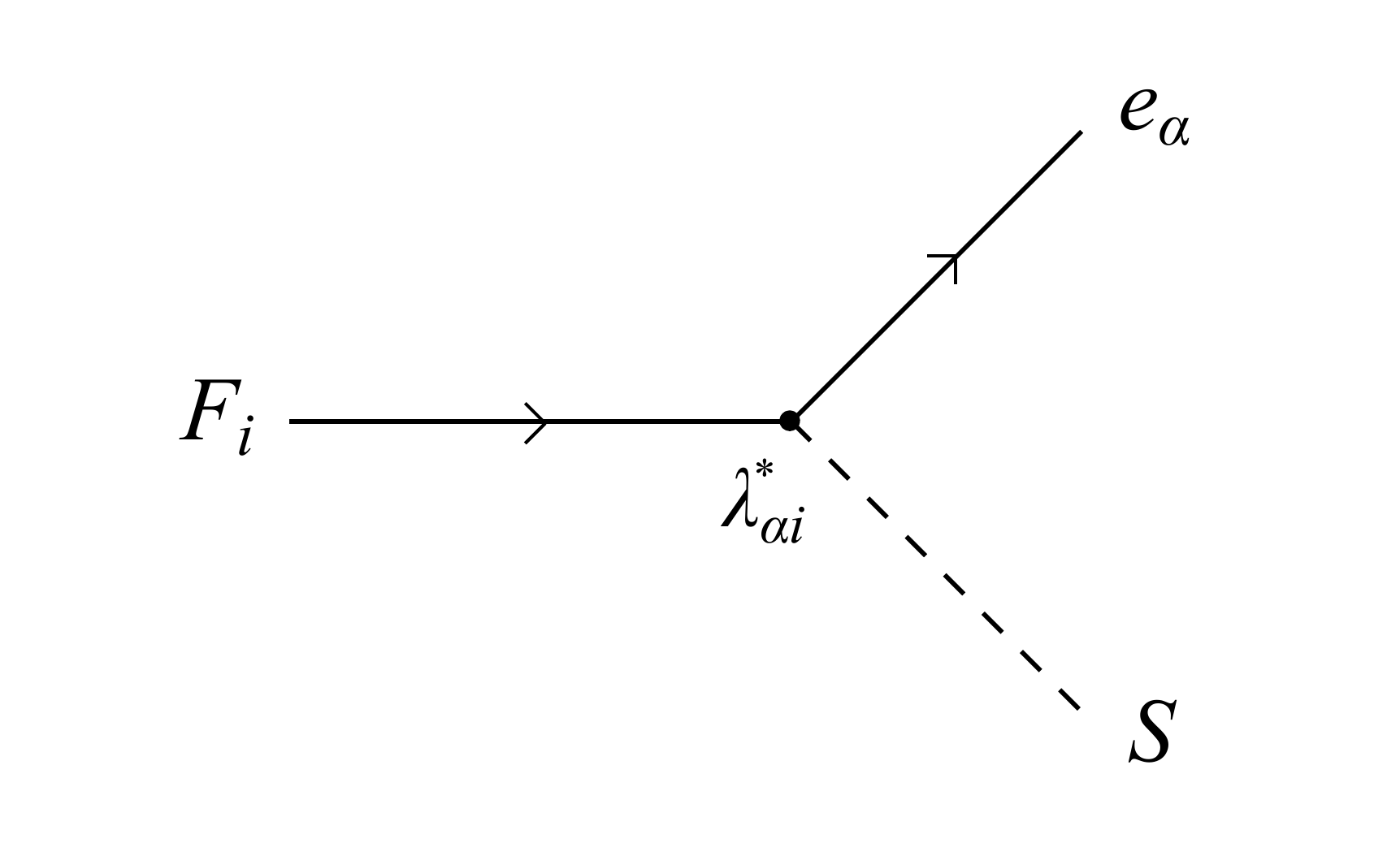}
  \caption{}
  \label{fig:tree-level}
\end{subfigure}\hspace*{\fill}
\begin{subfigure}{0.24\textwidth}
  \includegraphics[width=\linewidth]{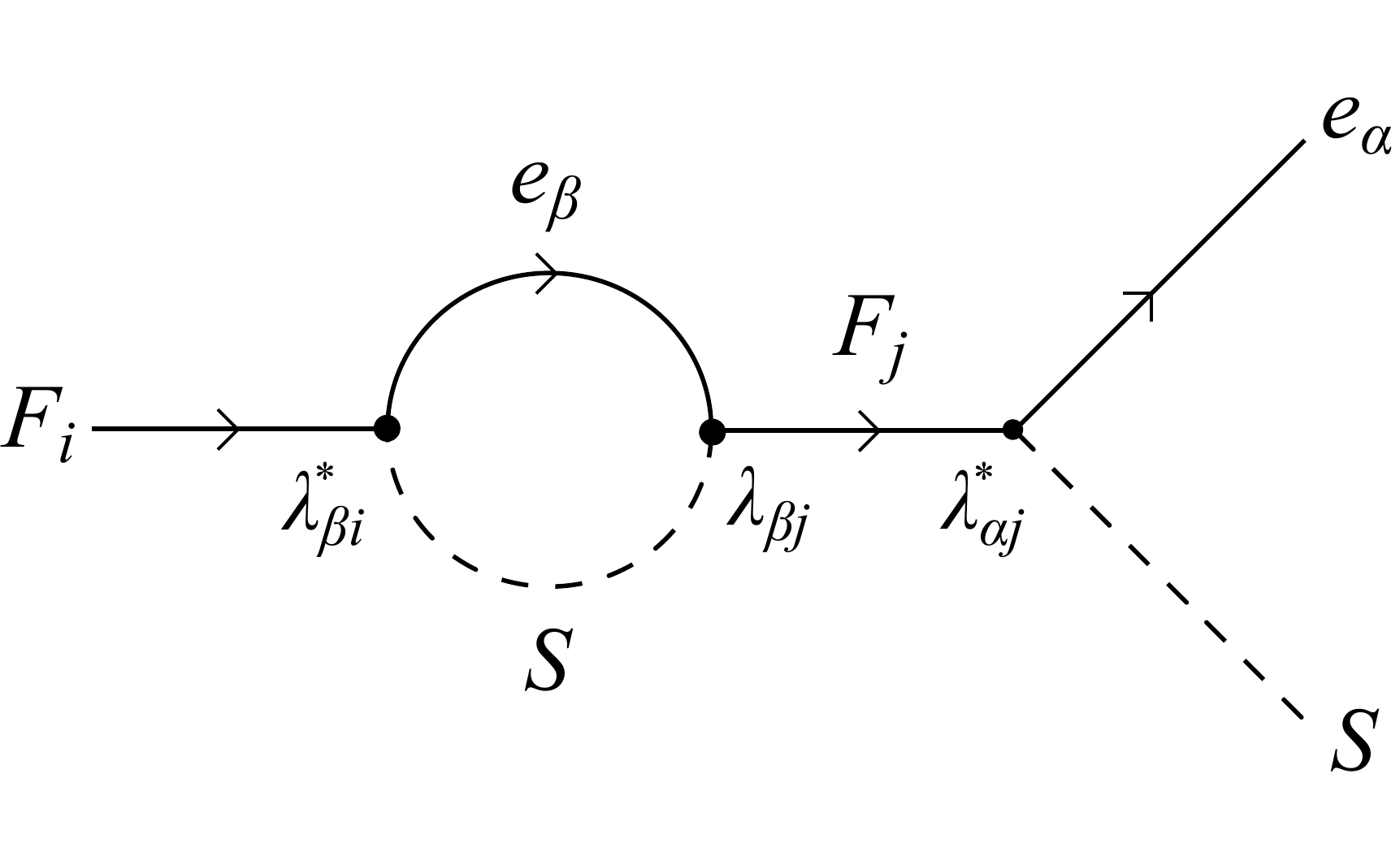}
  \caption{}
  \label{fig:loop}
\end{subfigure}\hspace*{\fill}
\begin{subfigure}{0.24\textwidth}
  \includegraphics[width=\linewidth]{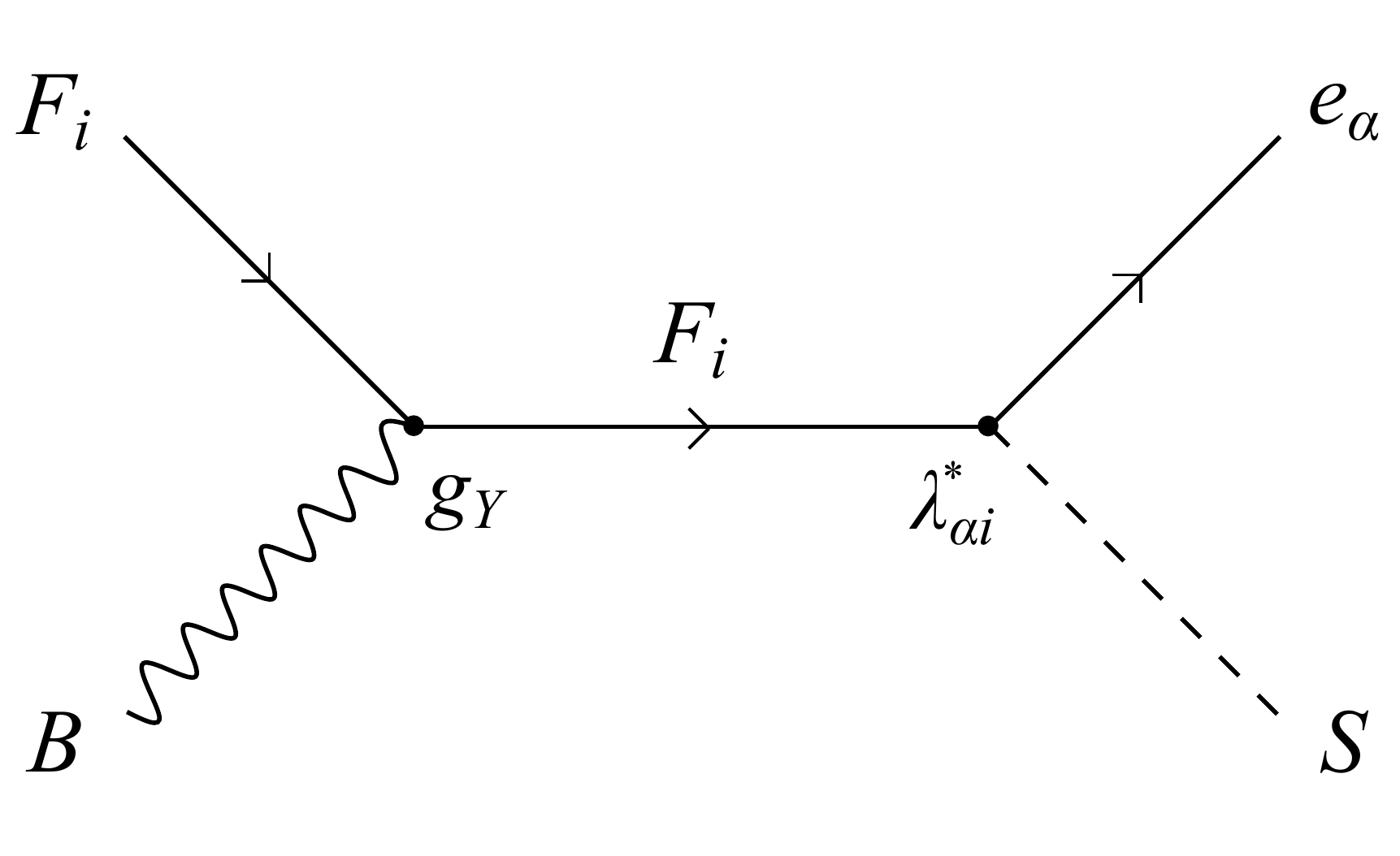}
  \caption{}
  \label{fig:gauge scattering 1-s}
\end{subfigure}\hspace*{\fill}
\begin{subfigure}{0.24\textwidth}
  \includegraphics[width=\linewidth]{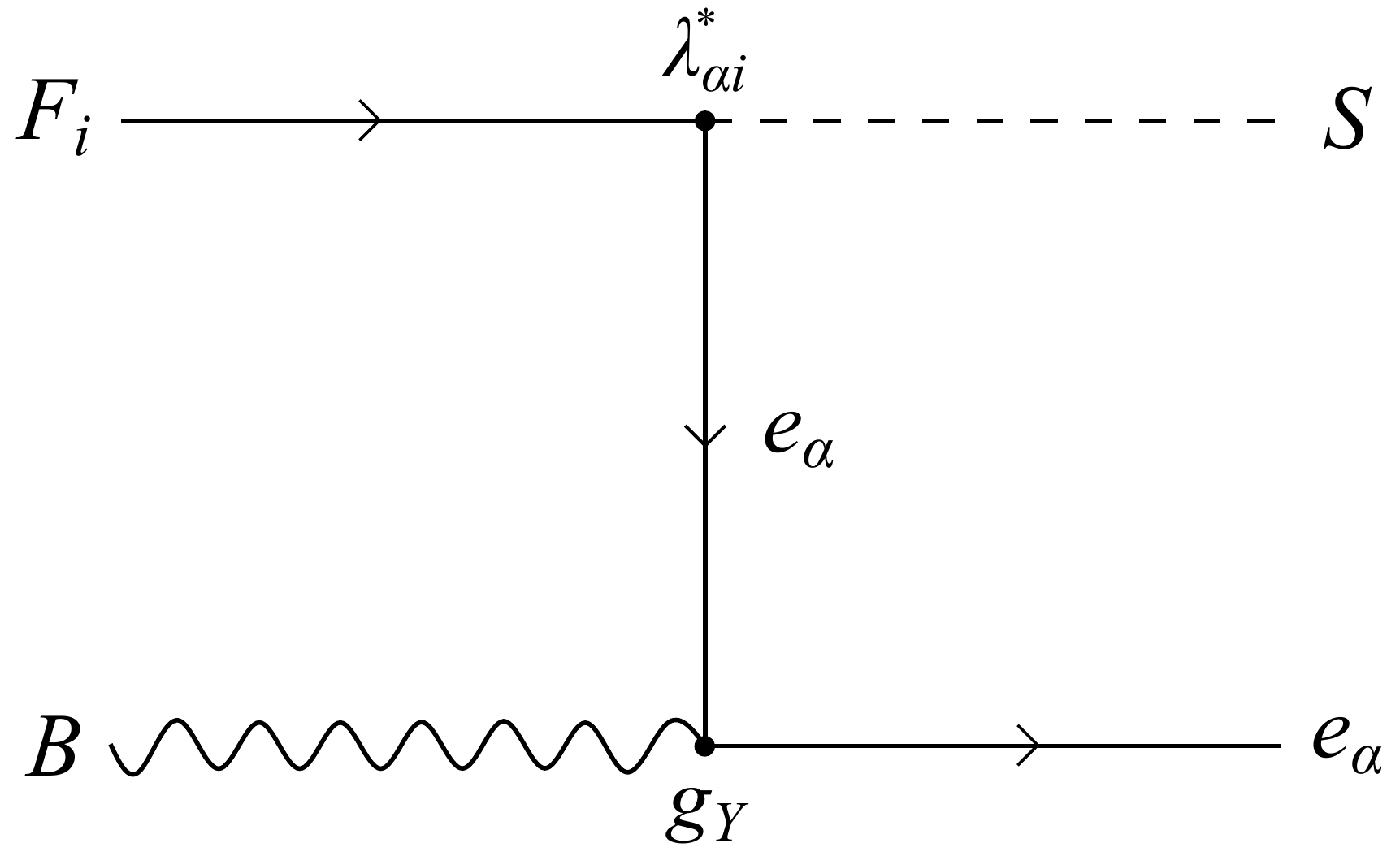}
  \caption{}
  \label{fig:gauge scattering 1-u}
\end{subfigure}

\medskip

\begin{subfigure}{0.24\textwidth}
  \includegraphics[width=\linewidth]{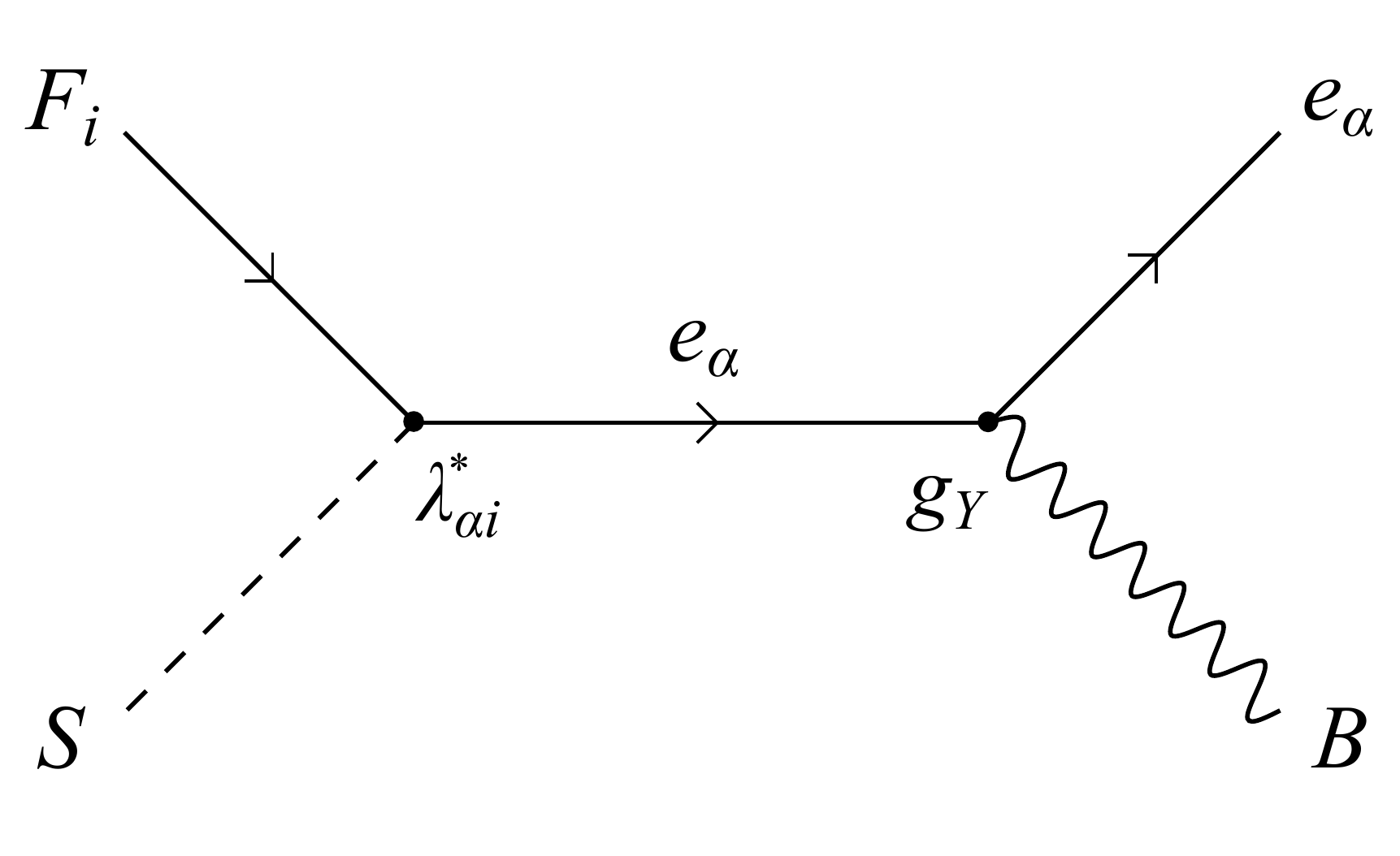}
  \caption{}
  \label{fig:gauge scattering 2-s}
\end{subfigure}\hspace*{\fill}
\begin{subfigure}{0.24\textwidth}
  \includegraphics[width=\linewidth]{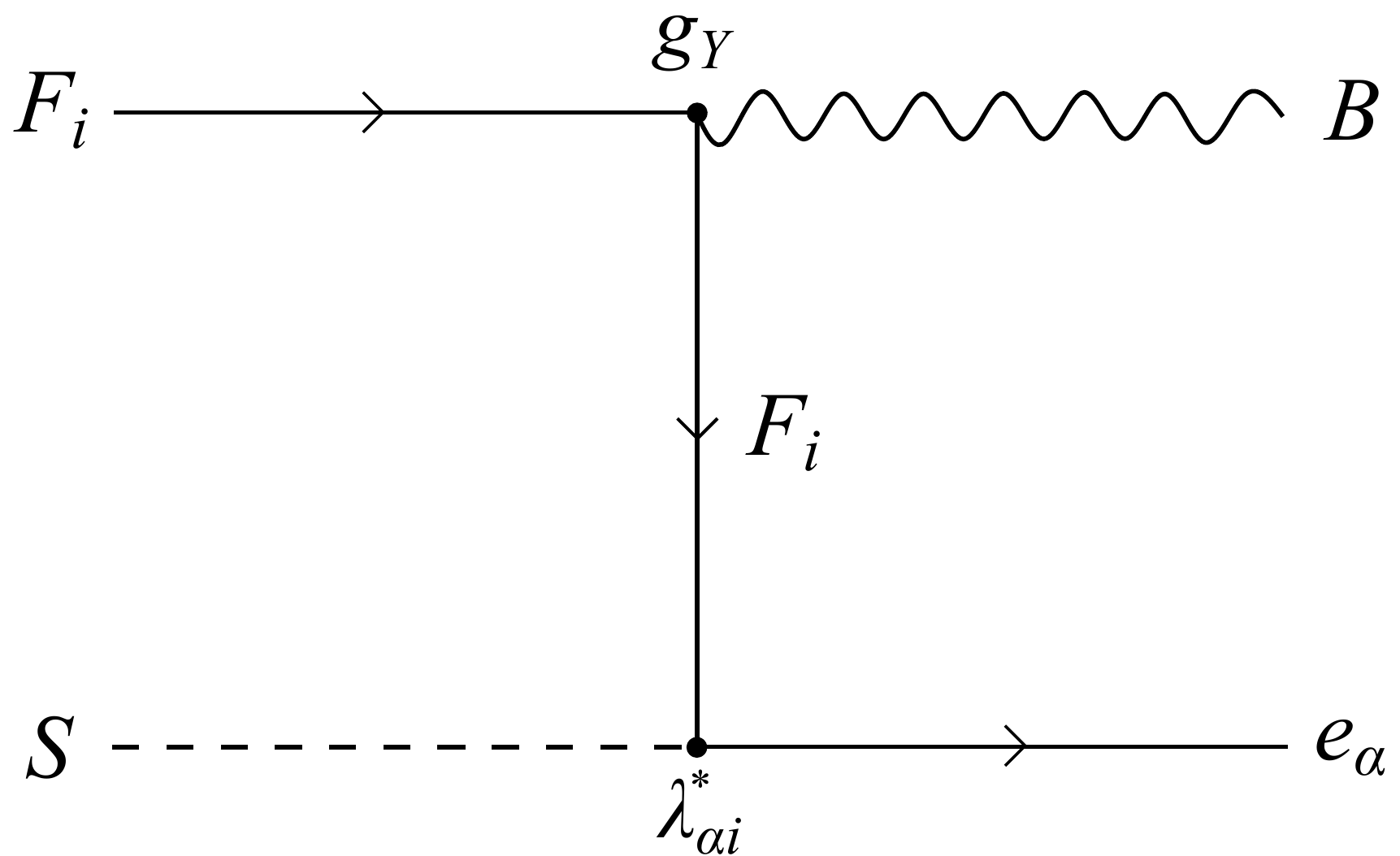}
  \caption{}
  \label{fig:gauge scattering 2-u}
\end{subfigure}\hspace*{\fill}
\begin{subfigure}{0.24\textwidth}
  \includegraphics[width=\linewidth]{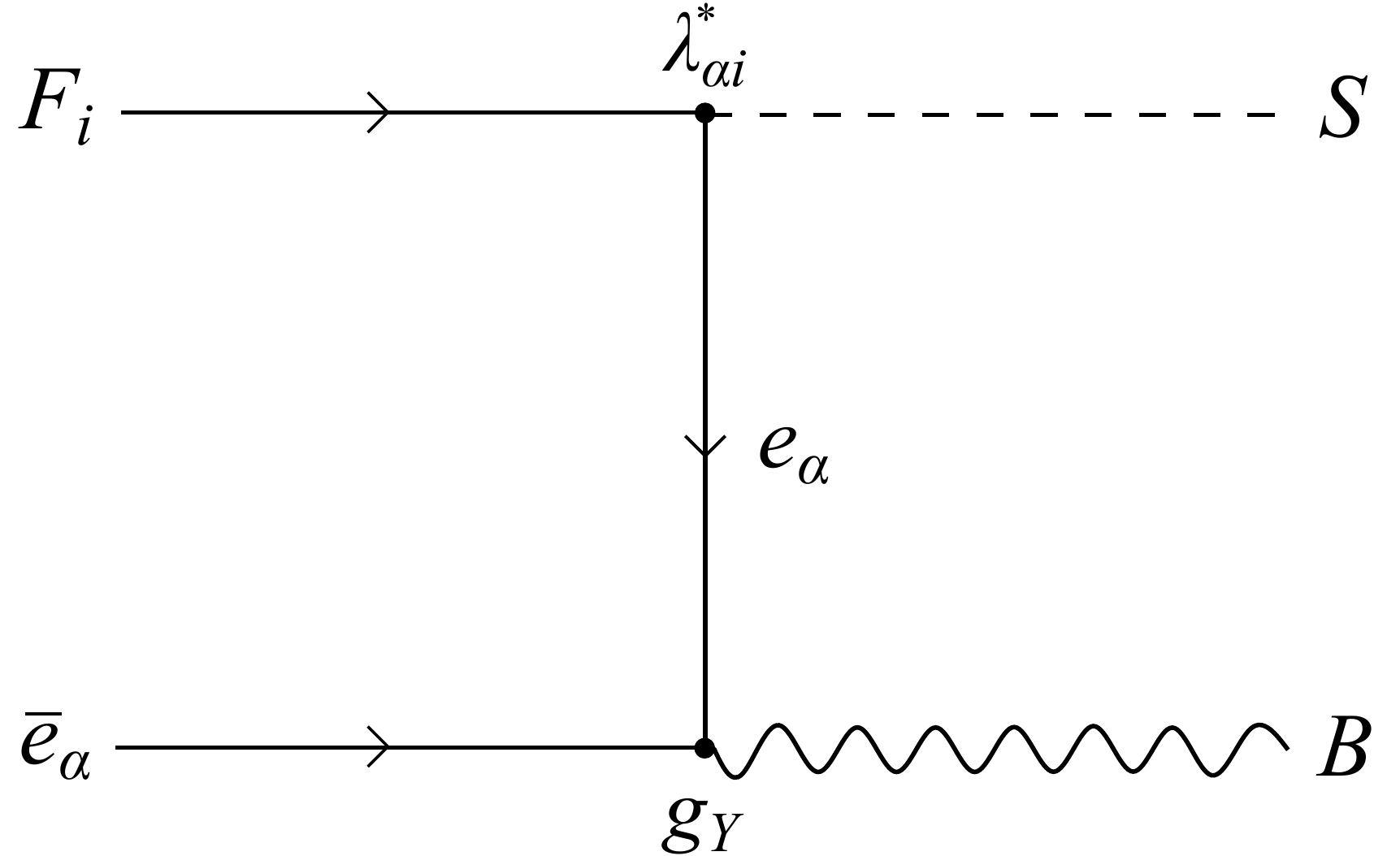}
  \caption{}
  \label{fig:gauge scattering 3-t}
\end{subfigure}\hspace*{\fill}
\begin{subfigure}{0.24\textwidth}
  \includegraphics[width=\linewidth]{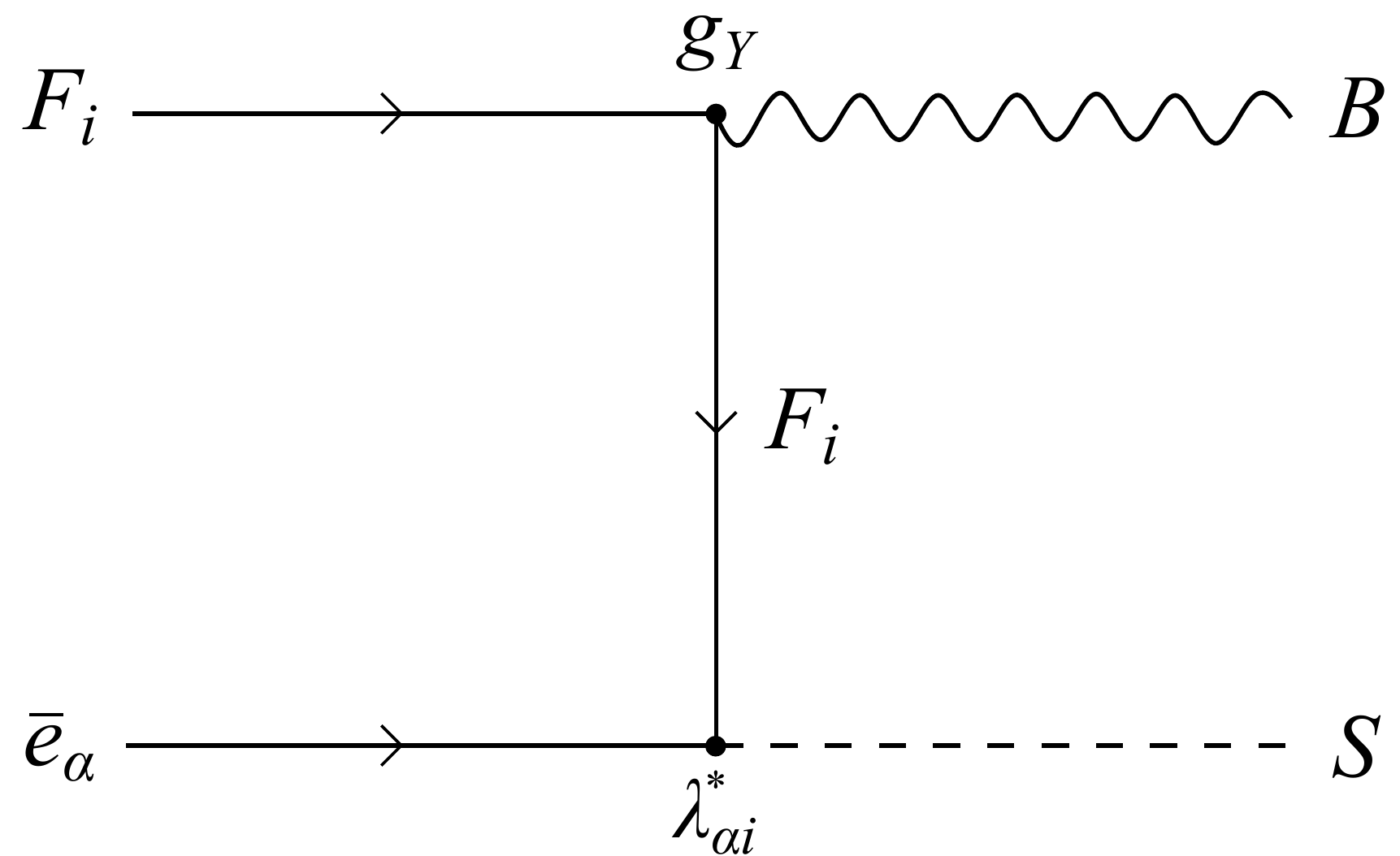}
  \caption{}
  \label{fig:gauge scattering 3-u}
\end{subfigure}
\caption{Feynman diagrams that contribute to the generation of the baryon asymmetry.}
\label{fig:Feynman diagrams}
\end{figure}

\noindent
where $\text{d}\Pi_k=\text{d}^3\textbf{p}_k/\left(2\pi\right)^32E_k$ is the elementary Lorentz invariant phase space volume of species $k$, $\left|\mathcal{M}\right|^2$ denotes the squared matrix element summed, but not averaged, over the internal degrees of freedom of the initial states, $K_1$ is the modified Bessel function of the second kind of order one and the distribution function of $F_i$ approximately follows the Maxwell-Boltzmann distribution, $f_{F_i}^{\text{eq}}= e^{-E_{F_i}/T}$. The corresponding thermally averaged equilibrium decay rate density $\Braket{\gamma_{F_i\rightarrow e_{\alpha}\,S}}$ reads

\begin{equation}
\Braket{\gamma_{F_i\rightarrow e_{\alpha}\,S}}\,\equiv\,\frac{\gamma_{F_i\rightarrow e_{\alpha}\,S}}{n_{F_i}^{\text{eq}}}\,=\,\frac{K_1\left(M_i/T\right)}{K_2\left(M_i/T\right)}\,\Gamma_{F_i\rightarrow e_{\alpha}\,S}\,\simeq\,\left\{\,
  	\begin{array}{@{}ll@{}}
    \frac{M_i}{2T}\,\Gamma_{F_i\rightarrow e_{\alpha}\,S},\quad &T\,\gg\,M_i
    \\\\
    \Gamma_{F_i\rightarrow e_{\alpha}\,S},\quad &T\,\ll\,M_i
	\end{array}
\right.
\end{equation}

\noindent
where $n_{F_i}^{\text{eq}}$ is the equilibrium number density of $F_i$ assuming zero chemical potential

\begin{equation}
n_{F_i}^{\text{eq}}\,\equiv\,g_{F_i}\int\frac{\text{d}^3\textbf{p}}{\left(2\pi\right)^3}\,f_{F_i}^{\text{eq}}\,=\,\frac{g_{F_i}}{2\pi^2}\,M_i^2\,T\,K_2\left(\frac{M_i}{T}\right)
\end{equation}
\\
\noindent
The time dilation factor $K_1\left(M_i/T\right)/K_2\left(M_i/T\right)$ implies that decays are inhibited at temperatures higher than the decaying state mass $M_i$.

The dominant $2\leftrightarrow 2$ scattering processes modifying the abundance of $S$ are those which involve the $U(1)_Y$ gauge boson as an external state, \textit{i.e.} $F_iB\leftrightarrow e_{\alpha}S$, $F_iS\leftrightarrow e_{\alpha}B$ and $F_i\bar{e}_{\alpha}\leftrightarrow SB$, which will be henceforth referred to as gauge scatterings. The corresponding matrix elements depend on the product of a feeble and a gauge coupling, whereas all other scattering processes involve higher powers of feeble couplings and are therefore subleading. All relevant Feynman diagrams are shown in Fig.~\ref{fig:Feynman diagrams}. 

\begin{figure}[t]
\centering
\includegraphics[width=0.7\linewidth]{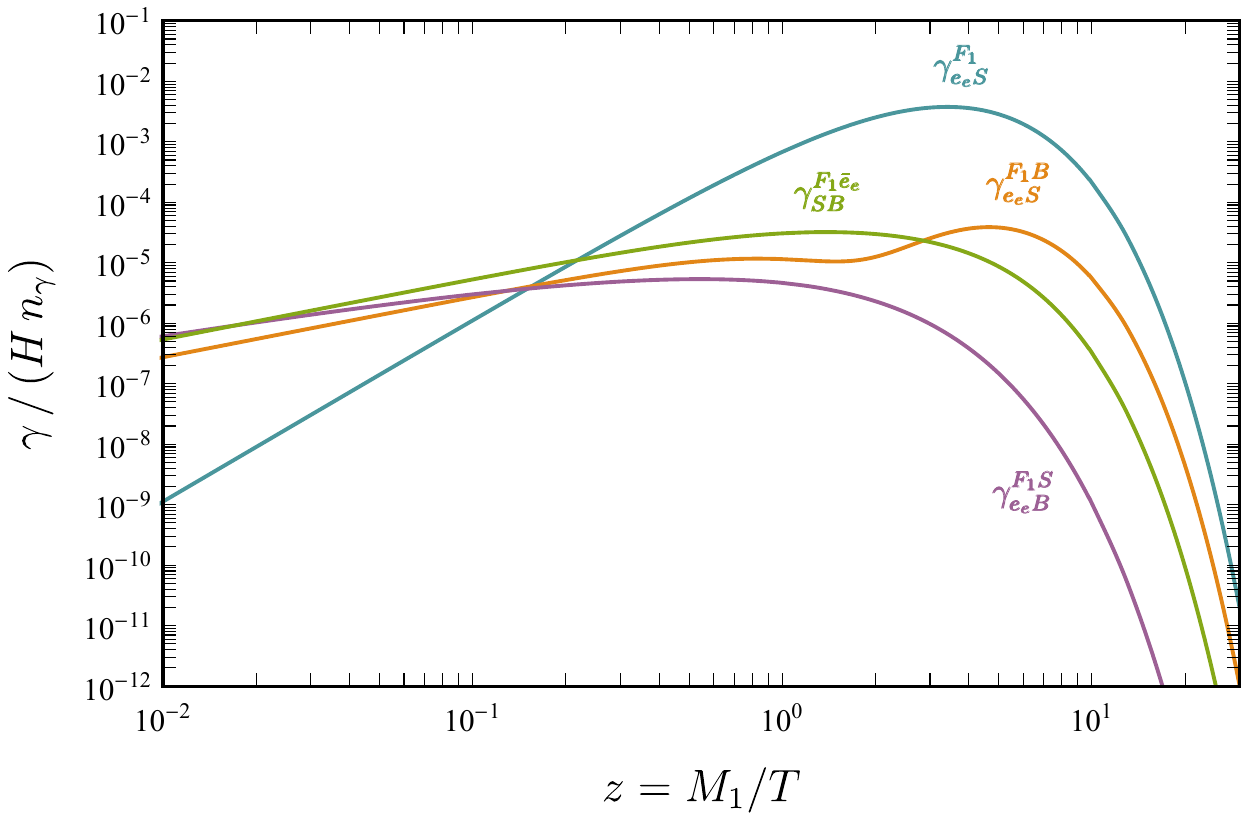}
\caption{Tree-level decay and gauge scattering rate densities involving $F_1$ and $e_e$ as external states, normalized to the Hubble parameter $H$ and the equilibrium number density of photons $n_{\gamma}$. The feeble coupling of the interactions is $\lambda_{e1}\simeq 2.15\times 10^{-8}$ and the densities have been evaluated at tree-level and in the limit $M_1\gg m_{e_{\alpha}}+m_S$. }
\label{fig:Rate Densities}
\end{figure}

The equilibrium interaction rate density of a generic scattering process $ab\rightarrow cd$ is \cite{Gondolo:1990dk}

\begin{align}
\gamma_{ab\rightarrow cd}\,&\equiv\,\int\text{d}\Pi_a\,\text{d}\Pi_b\,\text{d}\Pi_c\,\text{d}\Pi_d\,\left(2\pi\right)^4\,\delta^{(4)}\left(p_{a}+p_{b}-p_{c}-p_{d}\right)\,f^{\text{eq}}_{a}\,f^{\text{eq}}_{b}\,\left|\mathcal{M}\right|^2_{ab\rightarrow cd}\nonumber
\\
&=\,\frac{T}{512\pi^6}\,\int_{\tilde{s}_{\text{min}}}^{\infty}\,\text{d}\tilde{s}\,\frac{1}{\sqrt{\tilde{s}}}\,\left|\textbf{p}_{\text{in}}\right|\,\left|\textbf{p}_{\text{fin}}\right|\,K_1\left(\frac{\sqrt{\tilde{s}}}{T}\right)\int\,\text{d}\Omega\,\left|\mathcal{M}\right|^2_{ab\rightarrow cd}
\end{align}
\\
\noindent
where $\textbf{p}_{\text{in}}$, $\textbf{p}_{\text{fin}}$ and $\sqrt{\tilde{s}}$ are the initial and final momenta and the energy in the center-of-momentum frame respectively, with $\tilde{s}_{\text{min}}=\max\big\{\left(m_a+m_b\right)^2\,,\,\left(m_c+m_d\right)^2\big\}$. In Fig.~\ref{fig:Rate Densities}, we present the decay and gauge scattering rate densities involving $F_1$ and the right-handed electron $e_e$ as external states in terms of the dimensionless parameter $z\equiv M_1/T$. For the feeble coupling, we use the value $\lambda_{e1}= 2.145\times 10^{-8}$ and work at lowest order in perturbation theory and in the limit $M_1\gg m_{e_{\alpha}}+m_S$. In order to study their effect on the Boltzmann equations, it is convenient to normalize them to the Hubble parameter $H$ and to the number density of photons $n_{\gamma}=2\zeta(3)T^3/\pi^2$, where $\zeta$ is the Riemann zeta function \cite{Kolb:1990vq}. The integrals appearing in the various scattering rate densities have been computed with the Cuba numerical library \cite{Cuba_2005}.

The s-channel resonance that appears in the scattering $F_iB\leftrightarrow e_{\alpha}S$ has been regularized by the finite decay width of $F_i$. Also, the IR-induced resonances due to the exchange of massless SM leptons that appear in the u-channel of $F_iB\leftrightarrow e_{\alpha}S$ and in the t-channel of $F_i\bar{e}_{\alpha}\leftrightarrow SB$ have both been regulated by the thermal mass of the right-handed SM leptons $m_e^2\left(T\right)=g_Y^2T^2/8$ (see \cite{Cline_1994} and references therein). Note that, as thermal effects are irrelevant at the temperature of a few $\TeV$ that is of interest to us, we include the thermal mass only when it acts as a regulator of the IR-divergences. For a few examples in which finite temperature effects can become important \textit{cf e.g.} \cite{Baker:2017zwx,Dvorkin:2019zdi,Darme:2019wpd}.

\subsection{Out-of-equilibrium decays}\label{subsec:Out-of-Equilibrium Decays}

\noindent
As we already mentioned, one of the crucial ingredients of our setup is that all processes involving dark matter (and, for that matter, $CP$ violation) must never attain chemical equilibrium. In order to fulfill this condition, an upper bound on the magnitude of the freeze-in couplings can be obtained by requiring the total thermally averaged decay rate to be smaller than the Hubble expansion rate at the characteristic temperature $T=M_i$. Because of the feeble nature of the $\lambda_{\alpha i}$ couplings, successful baryogenesis requires a resonant enhancement of the asymmetry, which, in turn, implies that the masses of the two heavy fermions have to be very close to each other. Then, for $M_1\simeq M_2$ the out-of-equilibrium condition reads

\begin{equation}\label{eq:out-of-equilibrium condition}
\Big(\sum_{\alpha,i}\Braket{\gamma_{F_i\rightarrow e_{\alpha}\,S}}\,\lesssim\,H\,\Big)\big|_{T\,=\,M_1}
\end{equation}

\noindent
The Hubble parameter is given by

\begin{equation}\label{eq:Hubble parameter}
H\left(T\right)\,\simeq\,\frac{1.66\,\sqrt{g_{*\rho}}}{M_{Pl}}\,T^2
\end{equation}

\noindent
where $M_{Pl}\simeq 1.22\times 10^{19}\,\GeV$ is the Planck mass and $g_{*\rho}\simeq 106.75$ are the effective degrees of freedom related to the energy density. At leading order and in the limit $M_i\gg m_{e_{\alpha}}+m_{S}$, the out-of-equilibrium condition \eqref{eq:out-of-equilibrium condition} reads ($g_{F_i}=2$)

\begin{equation}
\sum_{\alpha,i}\Gamma_{F_i\rightarrow e_{\alpha}S}\,\lesssim\,2H\,\big|_{T\,=\,M_1}\;\Rightarrow\;\sum_{\alpha,i}\left|\lambda_{\alpha i}\right|^2\,\lesssim\,2.83\,\times\,10^{-13}\,\left(\frac{M_1}{\TeV}\right)
\end{equation}

\noindent
Thus, for heavy leptons at the $\TeV$ range, the couplings have to be smaller than $\sim 10^{-7}$ for the decays to proceed out-of-equilibrium.


\subsection{Freeze-in DM abundance}\label{subsec:Freeze-in DM abundance - Decays}

\noindent
The freeze-in DM abundance $Y_S$ that is produced from decays and $2\rightarrow 2$ scattering processes in our model follows the Boltzmann equation

\begin{align}\label{eq:Boltzmann DM full}
s\,\frac{\text{d}Y_S}{\text{d}t}\,&=\,\sum_{\alpha,i}\left\{F_i\,\leftrightarrow e_{\alpha}S\right\}\,+\,\sum_{\alpha,i}\left\{F_iB\leftrightarrow e_{\alpha}S\right\}\,-\,\sum_{\alpha,i}\left\{F_iS\leftrightarrow e_{\alpha}B\right\}\,+\,\sum_{\alpha,i}\left\{F_i\bar{e}_{\alpha}\leftrightarrow SB\right\}\nonumber
\\
&+\,2\sum_{i,j}\left\{F_i\bar{F}_j\leftrightarrow SS\right\}\,+\,2\sum_{\alpha,\beta}\left\{\bar{e}_{\alpha}e_{\beta}\leftrightarrow SS\right\} \ .
\end{align}

\noindent
In writing \eqref{eq:Boltzmann DM full}, we have used the notations

\begin{subequations}
\begin{alignat}{3}
\left\{a\,b\leftrightarrow c\,d\right\}\,&\equiv\,\left(a\,b\leftrightarrow c\,d\right)\,+\,\left(\bar{a}\,\bar{b}\leftrightarrow \bar{c}\,\bar{d}\right)
\\
\left[a\,b\leftrightarrow c\,d\right]\,&\equiv\,\left(a\,b\leftrightarrow c\,d\right)\,-\,\left(\bar{a}\,\bar{b}\leftrightarrow \bar{c}\,\bar{d}\right)
\\
\left(a\,b\leftrightarrow c\,d\right)\,&\equiv\,\int\text{d}\Pi_a\text{d}\Pi_b\text{d}\Pi_c\text{d}\Pi_d\left(2\pi\right)^4\delta^{(4)}\Big[\left|\mathcal{M}\right|^2_{ab\rightarrow cd}f_af_b\left(1\pm f_c\right)\left(1\pm f_d\right)\nonumber
\\
&\quad\qquad\qquad\qquad\qquad\qquad\qquad\qquad-\,\left|\mathcal{M}\right|^2_{cd\rightarrow ab}f_cf_d\left(1\pm f_a\right)\left(1\pm f_b\right)\Big]
\end{alignat}
\end{subequations}

\noindent
where $\delta^{\left(4\right)}$ is an abbreviation for $\delta^{\left(4\right)}\left(p_a+p_b-p_c-p_d\right)$, $f_k$ is the distribution function of species $k$ and $s$ denotes the entropy density. We will make the following assumptions:

\begin{itemize}
\item The initial DM abundance is zero. Combined with the feeble couplings, this allows us to ignore the inverse decays $e_{\alpha}S\rightarrow F_i$, \textit{i.e.} $f_S\simeq 0$. 

\item The DM production takes place during the radiation dominated era. At this epoch, time and temperature are related by $\dot{T}\simeq -HT$, which is valid for $\partial g_{*\rho}/\partial T\simeq 0$. Using this relation, we may switch variables and write

\begin{equation}
s\frac{\text{d}}{\text{d}t}Y_{S}\,=\,sHz\,\frac{\text{d}}{\text{d}z}Y_{S}
\end{equation} 

\noindent
The entropy density during the radiation dominated era is given by

\begin{equation}\label{eq:entropy density}
s\,=\,\frac{2\pi^2}{45}\,g_{*s}\,T^3
\end{equation}

\noindent
where $g_{*s}$ are the effective degrees of freedom with respect to the entropy density.

\item The distribution functions of the visible sector species obey the Maxwell-Boltzmann statistics, \textit{i.e.} we neglect Bose-enhancement and Pauli-blocking factors. Hence, we can write, in general,

\begin{equation}
\left(a\,b\leftrightarrow c\,d\right)\,=\,\gamma_{ab\rightarrow cd}\,\frac{Y_a}{Y_a^{\text{eq}}}\,\frac{Y_b}{Y_b^{\text{eq}}}\,-\,\gamma_{cd\rightarrow ab}\,\frac{Y_c}{Y_c^{\text{eq}}}\,\frac{Y_d}{Y_d^{\text{eq}}}
\end{equation}

\end{itemize}

Let us first focus on the heavy lepton decays $F_i\rightarrow e_{\alpha}S$ and the corresponding $CP$-conjugate processes, which provide the dominant contribution to DM production for $z\gtrsim 1$ \cite{Hall:2009bx}. Under the aforementioned assumptions, the Boltzmann equation for the DM abundance can be written as

\begin{align}\label{eq:Boltzmann equation DM decays}
sHz\frac{\text{d}}{\text{d}z}Y_{S}\,&\simeq\,\sum_{\alpha,i}\left\{F_i\,\leftrightarrow e_{\alpha}S\right\}\nonumber
\\
&=\,2\,\sum_{\alpha, i}\gamma^{F_i}_{e_{\alpha}S}\,\frac{Y_{F_i+\bar{F}_i}}{Y^{\text{eq}}_{F_i+\bar{F}_i}}\,+\,\mathcal{O}\left(\epsilon^2\right)\nonumber
\\
&=\,\frac{g_{F_1}}{\pi^2}\,M_1^2\,T\,K_1\left(\frac{M_1}{T}\right)\sum_{\alpha}\Gamma_{F_1\rightarrow e_{\alpha}S}\,+\,\frac{g_{F_2}}{\pi^2}\,M_2^2\,T\,K_1\left(\frac{M_2}{T}\right)\sum_{\alpha}\Gamma_{F_2\rightarrow e_{\alpha}S}
\end{align}

\noindent
where $\gamma^{F_i}_{e_{\alpha}S}$ is the equilibrium decay rate density at tree-level, $\epsilon$ denotes the $CP$ asymmetry and we have used $Y_{F_i+\bar{F}_i}\simeq Y_{F_i+\bar{F}_i}^{\text{eq}}$.\footnote{Gauge scatterings keep $F_i\left(\bar{F}_i\right)$ close to thermal equilibrium down to $z\sim 25$, when they eventually freeze-out. Since the baryon asymmetry is generated prior to their freeze-out (when sphalerons become inactive), this is a valid approximation.} If we consider $M_1\simeq M_2$ (resonant case) and substitute the decay width \eqref{eq:decay width}, the Hubble parameter \eqref{eq:Hubble parameter} and the entropy density \eqref{eq:entropy density} in Eq.~\eqref{eq:Boltzmann equation DM decays}, then at tree-level and in the limit $M_i\gg m_{e_{\alpha}}+m_{S}$ the DM abundance simplifies to

\begin{equation}\label{eq:DM abundance from decays}
Y_{S}\left(z\right)\,=\,\frac{45M_{Pl}}{32\times 1.66\,\pi^5\,g_{*s}\,\sqrt{g_{*\rho}}}\,\frac{\sum_{\alpha,i}\left|\lambda_{\alpha i}\right|^2}{M_1}\,\int_{z_{RH}}^{z}\text{d}z^{\prime}\,{z^{\prime}}^3\,K_1\left(z^{\prime}\right)
\end{equation}
\\
\noindent
where $z_{RH}\equiv M_1/T_{RH}$ and we have considered for simplicity that $\partial g_{*s}/\partial T\simeq 0$. In our analysis, $T_{RH}$ will be set to $10^{12}\,\GeV$. At the present day $T=T_0\simeq 2.73\,$K, so $T_0\ll M_1\ll T_{RH}$ and the integral contributes a factor of $3\pi/2$, yielding

\begin{equation}
Y_{S}\left(z_0\right)\,=\,\frac{135M_{Pl}}{64\times 1.66\,\pi^4\,g_{*s}\,\sqrt{g_{*\rho}}}\,\frac{\sum_{\alpha,i}\left|\lambda_{\alpha i}\right|^2}{M_1}
\end{equation}
\\
\noindent
The experimentally observed DM abundance is

\begin{equation}\label{eq:DM_Observed_Abundance}
Y_{S}\left(z_0\right)\,=\,\frac{\Omega_{DM}\,\rho_c}{m_{S}\,s_0}
\end{equation}

\noindent
where $\Omega_{\text{DM}}h^2= 0.1200\pm 0.0012$, $\rho_c\equiv 3H_0^2/8\pi\,G\simeq 10.537\,h^2\,\GeV m^{-3}$ is the critical density and $s_0\simeq 2.9\times 10^{9}\,m^{-3}$ is the entropy density at the present day \cite{2020}. If we ignore dark matter production through scattering processes, the DM mass $m_{S}$ can be related to the heavy lepton mass $M_{1}$ and to the feeble couplings as

\begin{equation}\label{eq:DM mass}
m_{S}\,=\,\frac{64\times 1.66\,\pi^4\,g_{*s}\,\sqrt{g_{*\rho}}\,\Omega_{DM}\,\rho_c}{135M_{Pl}\,s_0}\,\frac{M_1}{\sum_{\alpha,i}\left|\lambda_{\alpha i}\right|^2}
\end{equation}

\noindent
Lyman-$\alpha$ forest observations can be used in order to extract a lower bound on the DM mass, the exact value of which depends on the underlying mechanism of DM genesis. For DM candidates that freeze-out the current bound is $m_{DM}\gtrsim\left(1.9-5.3\right)\keV$ at $95\%$ C.L. \cite{garzilli2019warm,Palanque_Delabrouille_2020,Ir_i__2017}. This limit can be mapped onto the case of freeze-in-produced DM yielding $m_S\gtrsim\left(4-16\right)\keV$ \cite{Ballesteros_2021,Boulebnane_2018,DEramo:2020gpr,Decant:2021mhj}. This, in turn, imposes an upper limit on the feeble couplings

\begin{equation}\label{eq:Lyman-a bound}
\sum_{\alpha,i}\left|\lambda_{\alpha i}\right|^2\,\lesssim\,7.55\times 10^{-16}\left(\frac{M_1}{\TeV}\right)
\end{equation}
in order not to overclose the Universe, where we have used the more conservative bound $m_S\gtrsim 4\keV$. Thus, we see that the Lyman-$\alpha$ forest sets a more severe constraint than the out-of-equilibrium condition of Eq.~\eqref{eq:out-of-equilibrium condition}, forcing the feeble couplings to lie in the $10^{-8}$ range and below for heavy lepton masses at the $\TeV$ scale.

For a more rigorous treatment one should take into account the impact of scattering processes, which can modify the predicted DM abundance at $z\lesssim 1$, an epoch when scatterings are dominant. Including such scattering processes will be essential for the calculation of the baryon asymmetry presented in the next section. We focus only on scattering processes which involve gauge bosons as external states and ignore the subleading ones (second row of Eq.~\eqref{eq:Boltzmann DM full}). In this case, the Boltzmann equation takes the form

\begin{align}\label{eq:Boltzmann equation DM scatterings}
sHz\,\frac{\text{d}Y_S}{\text{d}z}\,&=\,\sum_{\alpha,i}\left\{F_i\,\leftrightarrow e_{\alpha}S\right\}\,+\,\sum_{\alpha,i}\left\{F_iB\leftrightarrow e_{\alpha}S\right\}\,-\,\sum_{\alpha,i}\left\{F_iS\leftrightarrow e_{\alpha}B\right\}\,+\,\sum_{\alpha,i}\left\{F_i\bar{e}_{\alpha}\leftrightarrow SB\right\}\nonumber
\\
&=\,2\,\sum_{\alpha,i}\Big(\gamma^{F_i}_{e_{\alpha}S}\,+\,\gamma^{F_iB}_{e_{\alpha}S}\,+\,\gamma^{F_i\bar{e}_{\alpha}}_{SB}\,+\,\gamma^{F_iS}_{e_{\alpha}B}\Big)\,+\,\mathcal{O}\left(\epsilon^2\right)
\end{align}

\begin{figure}[t]
\centering
\includegraphics[width=0.7\linewidth]{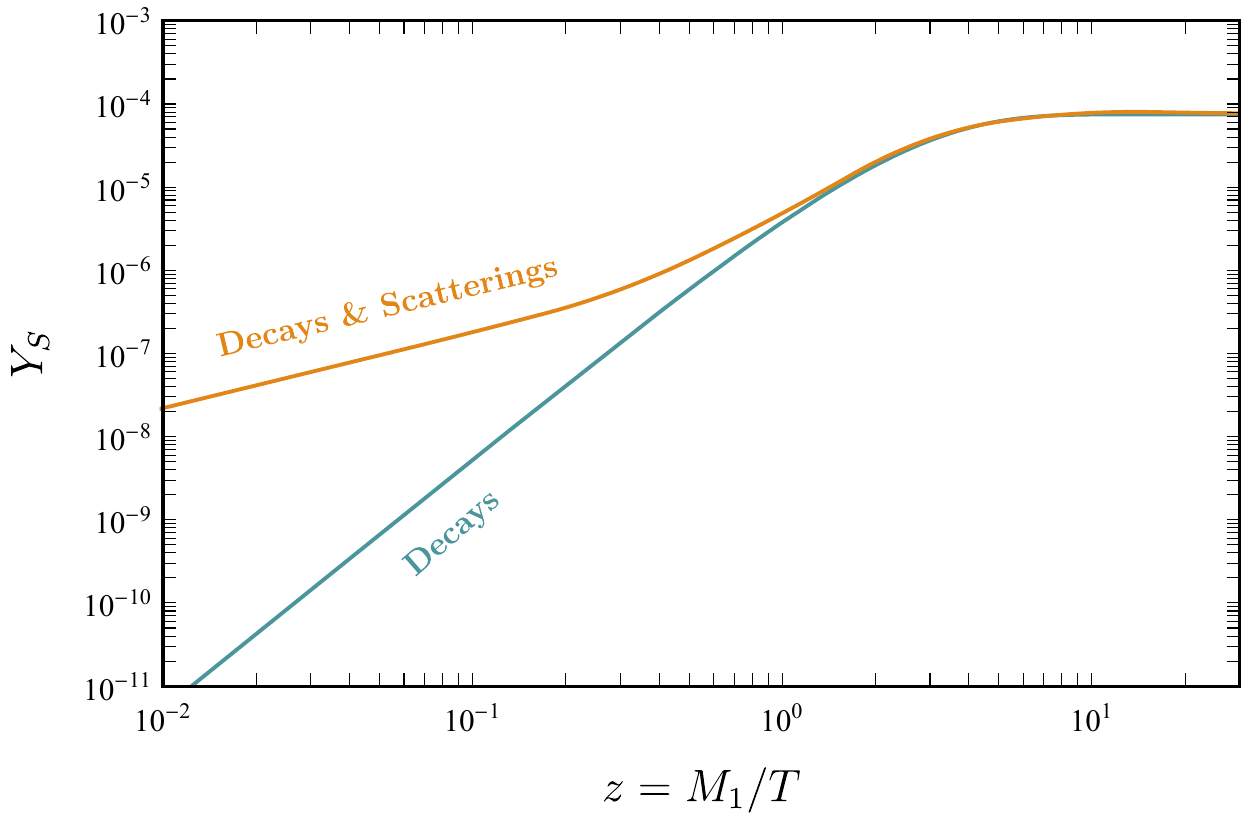}
\caption{DM abundance generated solely by decays, as well as by decays and scatterings for heavy leptons masses $M_1\simeq M_2 = 1.2\TeV$ and couplings $\sum_{\alpha,i}\left|\lambda_{\alpha i}\right|^2\simeq 6.25\times 10^{-16}$.}
\label{fig:DM Abundance}
\end{figure}

\noindent
where we have made use of the freeze-in approximation $f_S\simeq 0$ and considered $Y_{F_i+\bar{F}_i}\simeq Y_{F_i+\bar{F}_i}^{\text{eq}}$. In Figure \ref{fig:DM Abundance} we fix the masses of the heavy leptons $M_1\simeq M_2=1.2\TeV$ and the couplings $\sum_{\alpha,i}\left|\lambda_{\alpha i}\right|^2\simeq 6.25\times 10^{-16}$, and we compare the predicted DM abundance as estimated if we take into account only decays of the heavy leptons (lower, green line) with the result obtained through a full-blown numerical solution of Eq.~\eqref{eq:Boltzmann equation DM scatterings} (upper, yellow line). We observe that, as expected, the inclusion of the scattering processes does not drastically modify the predicted amount of dark matter in the Universe. Note that we have also cross-checked all of our results by implementing our model in {\tt FeynRules} \cite{Alloul:2013bka} and computing the predicted freeze-in DM abundance with {\tt CalcHEP/micrOMEGAs 5} \cite{Belyaev:2012qa, Belanger:2018ccd}. Given our findings, we conclude that the analytic estimate of the DM mass given by Eq.~\eqref{eq:DM mass} constitutes a reliable approximation. For the values of the physical parameters used in Figure \ref{fig:DM Abundance}, the value of the dark matter mass in order to reproduce the observed DM abundance in the Universe is $m_S\simeq 5.81\keV$, which is compatible with the Lyman-$\alpha$ forest bounds discussed previously.


\subsection{$CP$ asymmetry}\label{subsec:CP Asymmetry}

The $CP$ asymmetry generated through the decays of $F_{1,2}$ arises, at lowest order, due to the interference of the tree-level and 1-loop self-energy Feynman diagrams (wave-part contribution) as shown in Figures \ref{fig:tree-level} and \ref{fig:loop}. It can be defined in terms of the decay widths as

\begin{equation}\label{eq:CP Asymmetry def}
\epsilon_{\alpha\,i}\,\equiv\,\frac{\Gamma\left(F_i\rightarrow  e_{\alpha}S\right)\,-\,\Gamma\left(\bar{F}_i\rightarrow\bar{e}_{\alpha}S\right)}{\sum_{\alpha}\Gamma\left(F_i\rightarrow  e_{\alpha}S\right)\,+\,\Gamma\left(\bar{F}_i\rightarrow\bar{e}_{\alpha}S\right)}\, = \,\frac{\Gamma\left(F_i\rightarrow  e_{\alpha}S\right)\,-\,\Gamma\left(\bar{F}_i\rightarrow\bar{e}_{\alpha}S\right)}{2\,\Gamma_i}
\end{equation}

\noindent
where $\Gamma_i$ is the total decay width of $F_i$. If we separate the matrix element $\left.\mathcal{M}\right|_k$, with $k$ being the loop order $\left(k=0,1,\ldots\right)$, into a coupling constant part $c_k$ (denoting a collection of coupling constants) and an amplitude part $\mathcal{A}_k$, \textit{i.e.} $\left.\mathcal{M}\right|_k\equiv\sum_{i=0}^{k}\,c_i\mathcal{A}_i$, we can rewrite the $CP$ asymmetry as \cite{Davidson:2008bu}

\begin{equation}\label{eq:CP asymmetry 1}
\epsilon_{\alpha\,i}\,=\,-2\,\frac{\text{Im}\left\{c_0c_1^*\right\}}{\sum_{\alpha}\left|c_0\right|^2}\frac{\int\text{d}\Pi_{e_{\alpha}}\text{d}\Pi_{S}\left(2\pi\right)^4\delta^{(4)}\text{Im}\left\{\mathcal{A}_0\mathcal{A}_1^*\right\}}{\int\text{d}\Pi_{e_{\alpha}}\text{d}\Pi_{S}\left(2\pi\right)^4\delta^{(4)}\left|\mathcal{A}_0\right|^2}
\end{equation}

\noindent
The imaginary part of the amplitudes is related to the discontinuity of the corresponding graph as $2i\text{Im}\left\{\mathcal{A}_0\mathcal{A}_1^*\right\}=\text{Disc}\{\mathcal{A}_0\mathcal{A}_1^*\}$, which can be computed using the Cutkosky cutting rules. For a non-degenerate $F_i$ spectrum, $M_j-M_i\gg\Gamma_{j}$, we obtain \cite{PackageX}

\begin{align}\label{eq:CP asymmetry}
\epsilon_{\alpha i}\,=\,-\frac{1}{16\pi}\,\frac{1}{1\,-\,x_j}\,\frac{\text{Im}\Big\{\lambda_{\alpha i}^*\lambda_{\alpha j}\big[\lambda^{\dagger}\lambda\big]_{ji}\Big\}}{\left[\lambda^{\dagger}\lambda\right]_{ii}}
\end{align} 

\noindent
where $x_j\equiv M_j^2/M_i^2$ and we have used the standard notation $\left[\lambda^{\dagger}\lambda\right]_{ii}\equiv\left|\lambda_{ei}\right|^2+\left|\lambda_{\mu i}\right|^2+\left|\lambda_{\tau i}\right|^2$ and $\left[\lambda^{\dagger}\lambda\right]_{ji}\equiv\lambda_{ej}^* \lambda_{ei}+\lambda_{\mu j}^* \lambda_{\mu i}+\lambda_{\tau j}^* \lambda_{\tau i}$. This is half the result obtained in standard leptogenesis (see \textit{e.g.} Eq.~(5.13) in \cite{Davidson:2008bu}), because we deal with Dirac fermions where only a charged lepton propagates in the loop, whereas Majorana neutrinos can decay to both a charged lepton or a light neutrino accompanied by the Higgs field.

From $CPT$ invariance and unitarity, we know that the total decay width of a state and its $CP$-conjugate are equal

\begin{equation}\label{eq:total decay width}
\sum_{\alpha}\Gamma\left(F_i\rightarrow e_{\alpha}S\right)\,=\,\sum_{\alpha}\Gamma\left(\bar{F}_i\rightarrow\bar{e}_{\alpha}S\right)\,\equiv\,\Gamma_i
\end{equation}

\noindent
Hence, the $CP$ asymmetry vanishes when summed over the flavors, $\epsilon_i\,\equiv\,\sum_{\alpha}\epsilon_{\alpha\,i}\,=\,0$. This can be seen by summing Eq.~\eqref{eq:CP asymmetry} over flavors, a case in which the argument of the imaginary part becomes real and it vanishes identically:  
\begin{equation}
\sum_{\alpha}\text{Im}\Big\{\lambda_{\alpha i}^*\lambda_{\alpha j}\big[\lambda^{\dagger}\lambda\big]_{ji}\Big\}\,=\,\text{Im}\left\{\big[\lambda^{\dagger}\lambda\big]_{ij}\big[\lambda^{\dagger}\lambda\big]_{ji}\right\}\,=\,0
\end{equation}

\noindent
However, flavor effects \textit{can} lead to a non-vanishing baryon/lepton asymmetry, since washout processes are flavor dependent and, therefore, the lepton asymmetries in each flavor are washed out in a different way \cite{Abada_2006,Nardi_2006,Barbieri_2000,Blanchet_2007}.

Because of the feeble nature of the $\lambda_{\alpha i}$ couplings, a resonant enhancement of the $CP$ asymmetry is needed in order to generate a sufficiently large baryon asymmetry. This occurs when the mass difference between the heavy leptons is of the order of their decay widths and is related to the wave-part contribution to the $CP$ asymmetry (Figure \ref{fig:loop}). The resulting $CP$ asymmetry is given by

\begin{align}\label{eq:CP asymmetry regularized}
\epsilon_{\alpha i}\,=\,-\frac{1}{16\pi}\,\frac{1\,-\,x_j}{\left(1\,-\,x_j\right)^2\,+\,g_j^2}\,\frac{\text{Im}\Big\{\lambda_{\alpha i}^*\lambda_{\alpha j}\big[\lambda^{\dagger}\lambda\big]_{ji}\Big\}}{\left[\lambda^{\dagger}\lambda\right]_{ii}}
\end{align}

\noindent
where $g_j\equiv\Gamma_j/M_i$. In deriving \eqref{eq:CP asymmetry regularized}, we have regularized the divergence for small $x_j$ by applying the resummation procedure presented in \cite{Pilaftsis_1999}, with the regulator given by $M_1\Gamma_2$. Note that a more complete treatment of the resonance enhancement requires the employment of non-equilibrium techniques, which yield, in general, different regulators. Such an analysis has been performed \emph{e.g.} in \cite{Dev:2017wwc}, where the nearly degenerate states are considered to be out-of-equilibrium with the bath, in contrast to our model, where the $F_i$'s are kept close to equilibrium and the required deviation occurs for the DM states. Such an analysis is beyond the scope of this paper and is left for future work.

If we also express the coupling constants in terms of their magnitude and phase, $\lambda_{\alpha i}\,=\,\left|\lambda_{\alpha i}\right|e^{i\phi_{\alpha i}}$, Eq.~\eqref{eq:CP asymmetry regularized} can be rewritten as

\begin{align}
\epsilon_{\alpha i}\,&=\,-\frac{1}{16\pi}\,\frac{1\,-\,x_j}{\left(1\,-\,x_j\right)^2\,+\,g_j^2}\,\frac{\left|\lambda_{\alpha i}\right|\left|\lambda_{\alpha j}\right|}{\big[\lambda^{\dagger}\lambda\big]_{ii}}\,\sum_{\beta\neq\alpha}\left|\lambda_{\beta i}\right|\left|\lambda_{\beta j}\right|\sin\left(-\,\phi_{\alpha i}\,+\,\phi_{\alpha j}\,-\,\phi_{\beta j}\,+\,\phi_{\beta i}\right)\nonumber
\\
&\equiv\,-\frac{1}{16\pi}\,\frac{1\,-\,x_j}{\left(1\,-\,x_j\right)^2\,+\,g_j^2}\,\frac{\left|\lambda_{\alpha i}\right|\left|\lambda_{\alpha j}\right|}{\big[\lambda^{\dagger}\lambda\big]_{ii}}\,\sum_{\beta\neq\alpha}\left|\lambda_{\beta i}\right|\left|\lambda_{\beta j}\right|\,p^{ij}_{\alpha\beta}
\end{align}

\noindent
where $p^{ij}_{\alpha\beta}=-p^{ji}_{\alpha\beta}=-p^{ij}_{\beta\alpha}$. The resonance condition reads $M_2-M_1\sim \Gamma_2/2$ \cite{Pilaftsis_1999} and in this case the $CP$ asymmetry can be maximally enhanced to

\begin{subequations}
\begin{alignat}{2}
\epsilon_{\alpha 1}^{\text{res}}\,&=\,\frac{g_{F_1}}{2}\,\frac{\left|\lambda_{\alpha 1}\right|\left|\lambda_{\alpha 2}\right|}{\big[\lambda^{\dagger}\lambda\big]_{11}\big[\lambda^{\dagger}\lambda\big]_{22}}\,\sum_{\beta\neq\alpha}\left|\lambda_{\beta 1}\right|\left|\lambda_{\beta 2}\right|\,p^{12}_{\alpha\beta}
\\
\epsilon_{\alpha 2}^{\text{res}}\,&=\,g_{F_2}\,\frac{\left|\lambda_{\alpha 1}\right|\left|\lambda_{\alpha 2}\right|}{\big[\lambda^{\dagger}\lambda\big]_{11}^2\,+\,\big[\lambda^{\dagger}\lambda\big]_{22}^2}\,\sum_{\beta\neq\alpha}\left|\lambda_{\beta 1}\right|\left|\lambda_{\beta 2}\right|\,p^{12}_{\alpha\beta}
\end{alignat}
\end{subequations}
These are the expressions that we will be using in the numerical analysis that follows in order to compute $\epsilon_{\alpha i}$.


\subsection{Baryon asymmetry}

In the previous sections, we described how our model can satisfy two out of the three Sakharov conditions, namely, $C/CP$ violation and out-of-equilibrium dynamics. The last condition to be fulfilled in order to generate a baryon asymmetry is the violation of the baryon/lepton number. In the case of Majorana heavy states, a conserved lepton number cannot be consistently assigned in the presence of interaction and mass terms in the Lagrangian and, therefore, $L$ is violated. This is not the case in our model, where the heavy states are of Dirac nature. In this case, we may rely on sphaleron departure from equilibrium during the epoch that the lepton asymmetry is generated, along the lines described in \cite{Gonzalez:2009}. 

In the model we propose, the heavy leptons $F_i$ carry the same lepton number as the SM leptons and the total lepton asymmetry is $Y_L=Y_{L_{\text{SM}}}+Y_{L_F}$, where $Y_{L_{\text{SM}}}\equiv \sum_{\alpha}Y_{L_{\alpha}}$ and $Y_{L_F}=\sum_i Y_{L_{F_i}}$. All processes conserve the combination $Y_{B-L}\equiv Y_B-Y_{L_{\text{SM}}}-Y_{L_F}$, \textit{i.e.} $\text{d}Y_{B-L}/\text{d}z=0$. We also assume that the Universe is initially totally symmetric, $Y_{B-L_{\text{SM}}}|_{z_{RH}}=Y_{L_F}|_{z_{RH}}=0$ and, therefore, at any $z$ it holds $Y_{B-L_{\text{SM}}}=Y_{L_F}$. 

Since sphalerons are insensitive to the lepton asymmetry $Y_{L_F}$, as $F_i$ are $SU(3)_{\text{c}}\times SU(2)_{\LH}$-singlets, they affect only the non-zero lepton asymmetry stored in the SM lepton sector $Y_{L_{\text{SM}}}$. In particular, they convert it to a baryon asymmetry by imposing certain relations among the chemical potentials of the various species (see the Appendix). Once sphalerons depart from equilibrium, which occurs at $T_{sph}= 131.7\pm 2.3\GeV$ \cite{D_Onofrio_2014}, the baryon and lepton numbers are separately conserved. When the heavy leptons decay away, the total baryon asymmetry, being proportional to $Y_{B-L_{\text{SM}}}$, vanishes. However, if sphalerons become inactive during the decay epoch of the heavy leptons, then the baryon asymmetry freezes at $Y_B\propto Y_{B-L_{\text{SM}}}|_{T_{sph}}$, which, in general, is not null.

Taking into consideration all the decay and $2\rightarrow 2$ scattering processes, the full Boltzmann equations of the asymmetries read

\begin{align}\label{eq:Boltzmann Asymmetry 0}
-sHz\frac{\text{d}Y_{\Delta F_i}}{\text{d}z}\,&=\,\sum_{\alpha}\left[F_i\,\leftrightarrow e_{\alpha}S\right]\,+\,\sum_{\alpha}\left[F_iB\leftrightarrow e_{\alpha}S\right]\,+\,\sum_{\alpha}\left[F_iS\leftrightarrow e_{\alpha}B\right]\,+\,\sum_{\alpha}\left[F_i\bar{e}_{\alpha}\leftrightarrow SB\right]\nonumber
\\
&+\,\sum_{\alpha,\beta,j}\left[F_i\bar{e}_{\alpha}\leftrightarrow\bar{F}_je_{\beta}\right]\,+\,\sum_{\alpha,\beta}\left[F_i\bar{e}_{\alpha}\leftrightarrow F_j\bar{e}_{\beta}\right]\,+\,\sum_{\alpha,\beta}\left[F_ie_{\beta}\leftrightarrow F_je_{\alpha}\right]\nonumber
\\
&+\,\sum_{\alpha,\beta}\left[F_i\bar{F}_{j\neq i}\leftrightarrow e_{\alpha}\bar{e}_{\beta}\right]\,+\,\left[F_i\bar{F}_{j\neq i}\leftrightarrow SS\right]\,+\,\sum_{\alpha,\beta,j}\left[F_iF_j\leftrightarrow e_{\alpha}e_{\beta}\right]\nonumber
\\
&+\,\left[F_iS\leftrightarrow F_{j\neq i}S\right]
\end{align}

\begin{align}\label{eq:Boltzmann Asymmetry 00}
-sHz\frac{\text{d}Y_{\Delta_{\alpha}}}{\text{d}z}\,&=\,\sum_{i}\left[F_i\,\leftrightarrow e_{\alpha}S\right]\,+\,\sum_{i}\left[F_iB\leftrightarrow e_{\alpha}S\right]\,+\,\sum_{i}\left[F_iS\leftrightarrow e_{\alpha}B\right]\,+\,\sum_{i}\left[F_i\bar{e}_{\alpha}\leftrightarrow SB\right]\nonumber
\\
&+\,\sum_{i,j,\beta}\left[F_i\bar{e}_{\alpha}\leftrightarrow \bar{F}_je_{\beta}\right]\,+\,\sum_{i,j,\beta\neq\alpha}\left[F_i\bar{e}_{\alpha}\leftrightarrow F_j\bar{e}_{\beta}\right]\,+\,\sum_{i,j,\beta\neq\alpha}\left[F_ie_{\beta}\leftrightarrow F_je_{\alpha}\right]\nonumber
\\
&+\,\sum_{i,j,\beta\neq\alpha}\left[F_i\bar{F}_j\leftrightarrow e_{\alpha}\bar{e}_{\beta}\right]\,+\,\sum_{\beta\neq\alpha}\left[\bar{e}_{\alpha}e_{\beta}\leftrightarrow SS\right]\,+\,\sum_{i,j,\beta}\left[F_iF_j\leftrightarrow e_{\alpha}e_{\beta}\right]\nonumber
\\
&+\sum_{\beta\neq\alpha}\big[e_{\beta}S\leftrightarrow e_{\alpha}S\big]^{\prime}
\end{align}

\noindent
where $Y_{\Delta F_i}\equiv Y_{F_i}-Y_{\bar{F}_i}$ and $Y_{\Delta_{\alpha}}\equiv Y_B/3-Y_{L_{\text{SM}{\alpha}}}$. The primed term indicates that one has to subtract the contribution due to the on-shell propagation of $F_i$, usually referred to as real intermediate state subtraction (RISS), which is already taken into account by the successive decays $e_{\alpha}S\leftrightarrow F_i\leftrightarrow e_{\alpha}S$. 

The various terms in the Boltzmann equations can be expressed in terms of the $CP$ asymmetry $\epsilon_{\alpha i}$, the tree-level rate densities and the asymmetric abundances. As is typically done, we linearize in the SM chemical potentials \cite{Kolb:1979qa}

\begin{equation}
\frac{Y_{e_{\alpha}\left(\bar{e}_{\alpha}\right)}}{Y_{e_{\alpha}}^{\text{eq}}}\,\equiv\,e^{\pm\mu_{e_{\alpha}}/T}\,\simeq\,1\,\pm\,\frac{\mu_{e_{\alpha}}}{T}\,=\,1\,\pm\,\frac{Y_{\Delta e_{\alpha}}}{2Y_{\gamma}}    
\end{equation}

\noindent
All non-gauge interactions are subleading in comparison to the gauge interactions and can be safely discarded. We include only the $CP$-violating part of the RISS term, which ensures that the source term (proportional to $\epsilon_{\alpha i}$) takes the correct form, \textit{i.e.} it vanishes when all species are in chemical equilibrium. The Boltzmann equations can be rewritten as\footnote{In deriving Eq.~\eqref{eq:Boltzmann equation DM scatterings}, we dropped dark matter annihilation processes altogether. In baryo- and leptogenesis, inverse reactions can be crucial and need to be taken into account. In order to do so in an efficient manner, in Eqs.~\eqref{eq:Boltzmann Asymmetry 1} and \eqref{eq:Boltzmann Asymmetry 2} we will approximate $f_S \simeq f_S^{\rm eq} \frac{Y_S}{Y_S^{\rm eq}}$ -- a relation which is rigorously applicable to bath particles. We have checked that -- as expected, since we have restricted ourselves to regions of the parameter space in which the condition of Eq.~\eqref{eq:out-of-equilibrium condition} is satisfied -- this is, indeed, a good approximation which leads to only a small ($\sim 2\%$) reduction of the predicted dark matter abundance for large values of $z$.}

\begin{align}
-sHz\frac{\text{d}Y_{\Delta F_i}}{\text{d}z}\,&=\,\sum_{\alpha}\Big(y_{F_i}\,-\,\frac{Y_S}{Y_S^{\text{eq}}}\,y_{e_{\alpha}}\Big)\,\Big(\gamma^{F_i}_{e_{\alpha}S}\,+\,\gamma^{F_iB}_{e_{\alpha}S}\Big)\,+\,\sum_{\alpha}\Big(\frac{Y_S}{Y_S^{\text{eq}}}\,y_{F_i}\,-\,y_{e_{\alpha}}\Big)\,\gamma^{F_iS}_{e_{\alpha}B}\nonumber
\\
&+\,\sum_{\alpha}\Big(y_{F_i}\,-\,y_{e_{\alpha}}\Big)\,\gamma^{F_i\bar{e}_{\alpha}}_{SB}\label{eq:Boltzmann Asymmetry 1}
\\\nonumber
\\
-sHz\frac{\text{d}Y_{\Delta_{\alpha}}}{\text{d}z}\,&=\,2\,\sum_i\epsilon_{\alpha i}\,\Big[\,\Big(1\,-\,\frac{Y_S}{Y_S^{\text{eq}}}\Big)\,\sum_{\rho}\Big(\gamma^{F_i}_{e_{\rho}S}\,+\,\gamma^{F_iB}_{e_{\rho}S}\,+\,\gamma^{F_i\bar{e}_{\rho}}_{SB}\,-\,\gamma^{F_iS}_{e_{\rho}B}\Big)\Big]\nonumber
\\
&+\,\sum_i\Big[y_{F_i}\,-\,\frac{Y_S}{Y_S^{\text{eq}}}\,\sum_{\beta}\left(\text{B}^{F_i}_{e_{\beta}S}\,y_{e_{\beta}}\right)\Big]\,\gamma^{F_i}_{e_{\alpha}S}\,+\,\sum_i\Big(y_{F_i}\,-\,\frac{Y_S}{Y^{\text{eq}}_S}\,y_{e_{\alpha}}\Big)\,\gamma^{F_iB}_{e_{\alpha}S}\nonumber
\\
&+\,\sum_i\Big(\frac{Y_S}{Y_S^{\text{eq}}}\,y_{F_i}\,-\,y_{e_{\alpha}}\Big)\,\gamma^{F_iS}_{e_{\alpha}B}\,+\,\sum_i\Big(y_{F_i}\,-\,y_{e_{\alpha}}\Big)\,\gamma^{F_i\bar{e}_{\alpha}}_{SB}\label{eq:Boltzmann Asymmetry 2}
\end{align}

\noindent
where $y_{F_i}\equiv Y_{\Delta F_i}/Y_{F_i}^{\text{eq}}$, $y_{e_{\alpha}}\equiv Y_{\Delta e_{\alpha}}/Y_{\gamma}$, $\text{B}^{F_i}_{e_{\beta}S}$ denotes the branching ratio of the decay $F_i\rightarrow e_{\beta}S$ and we have used $Y_{F_i+\bar{F}_i}\simeq Y_{F_i+\bar{F}_i}^{\text{eq}}$. In deriving the equations above, we have taken into account that the $CP$ asymmetry in scattering processes stemming from self-energy diagrams (Figure \ref{fig:loop}) is always equal to the $CP$ asymmetry from decays \cite{Nardi_2007}. As described in \cite{Davidson:2008bu} one must also include the contributions from the RISS $2\rightarrow 3$ scatterings involving gauge bosons, in order to obtain the correct form for the scattering source terms. These processes can, however, be neglected as subleading with regards to their impact on washout. Note that the source term in the Boltzmann equation of the asymmetry in $F_i$ vanishes as a consequence of $CPT$ and unitarity, as explained in Section \ref{subsec:CP Asymmetry}. Lastly, these equations are not totally independent, as they are related through the conservation of $Y_{B-L}$, resulting in $\sum_{i}Y_{\Delta F_i}=\sum_{\alpha}Y_{\Delta\alpha}$.

One can express the asymmetries in each SM flavor $y_{e_{\alpha}}$ in terms of $Y_{\Delta\alpha}$, solving a system of algebraic equations which relate the chemical potentials and abundances in equilibrium of the various species (see the Appendix). For the temperatures of interest we find

\begin{equation}
\begin{pmatrix}
y_{e_e} \\ 
y_{e_{\mu}} \\ 
y_{e_{\tau}}
\end{pmatrix}
\,=\,
-\frac{1}{711}\,
\left[\begin{array}{ccc}
230 & -7 & -7 \\
-7 & 230 & -7 \\
-7 & -7 & 230
\end{array}\right]\,
\begin{pmatrix}
Y_{\Delta_e} \\ 
Y_{\Delta_{\mu}} \\ 
Y_{\Delta_{\tau}}
\end{pmatrix}\frac{1}{Y_{\gamma}}
\end{equation}
\\
\noindent
In the same way, we have calculated the amount of the SM lepton asymmetry converted to baryon asymmetry by the sphaleron transitions to be

\begin{equation}
Y_{B}\,=\,\frac{22}{79}\,\sum_{\alpha}Y_{\Delta\alpha}    
\end{equation}

\begin{figure}[t]
\centering
\includegraphics[width=0.75\linewidth]{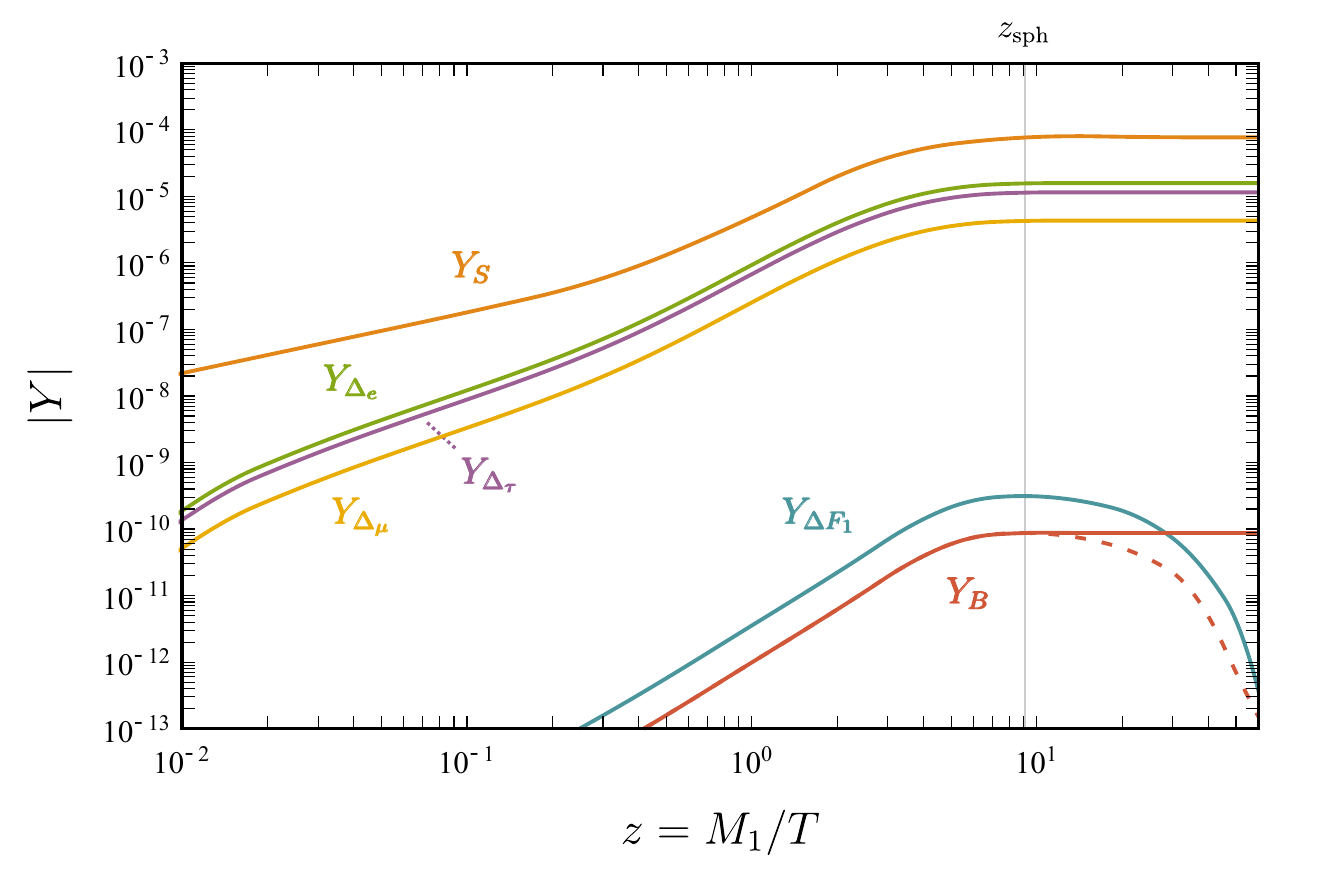}
\caption{The generated DM abundance $Y_S$, the asymmetries $Y_{\Delta\alpha}$, $Y_{\Delta F_1}$ and the baryon asymmetry $Y_B$ as a function of the dimensionless parameter $z=M_1/T$, for the heavy leptons masses and feeble couplings given in Eq.~\eqref{eq:parameter values}. The dashed line illustrates the evolution that $Y_B$ would follow in the absence of sphaleron decoupling, while the vertical line denotes the temperature at which sphalerons depart from equilibrium.}
\label{fig:Asymmetries}
\end{figure}

The system of the five coupled Boltzmann equations of the asymmetries \eqref{eq:Boltzmann Asymmetry 1} and \eqref{eq:Boltzmann Asymmetry 2} has been solved numerically, using the DM abundance generated by decays and scattering processes of Eq.~\eqref{eq:Boltzmann equation DM scatterings}. We confirm that the proposed model can indeed generate the observed baryon asymmetry $Y_{\text{B}}= 8.71\pm 0.06\times 10^{-11}$ \cite{2020} in the resonant regime. As an illustration, we present in Figure \ref{fig:Asymmetries} an explicit example, using the following values of the parameters:

\begin{equation}\label{eq:parameter values}
  \begin{split}
    M_1\,&=\,1.2\TeV,\\
    \left|\lambda_{e1}\right|\,&=\,2.145\times 10^{-8},\\
    \left|\lambda_{\mu 1}\right|\,&=\,\left|\lambda_{\tau 1}\right|\,=\,9\times 10^{-9},\\
    p^{12}_{e\mu}\,&=\,p^{12}_{e\tau}\,=\,p^{12}_{\mu\tau}\,=\,1
  \end{split}
\qquad
  \begin{split}
    M_2\,&=\,M_1\,+\,\Gamma_2/2\\
    \left|\lambda_{e2}\right|\,&=\,\left|\lambda_{\tau 2}\right|\,=\,9\times 10^{-10}\\
    \left|\lambda_{\mu 2}\right|\,&=\,8\times 10^{-10}\\
    &
  \end{split}
\end{equation}
\\
\noindent
The corresponding resonant $CP$ asymmetries $\epsilon_{\alpha i}$ turn out to be: $\epsilon_{e1}\simeq 2.1\times 10^{-1},\,\epsilon_{\mu 1}\simeq -5.7\times 10^{-2},\,\epsilon_{\tau 1}\simeq -1.53\times 10^{-1}$ and $\epsilon_{e2}\simeq 1.53\times 10^{-3},\,\epsilon_{\mu 2}\simeq -4.2\times 10^{-4},\,\epsilon_{\tau 2}\simeq -1.11\times 10^{-3}$. We have explicitly verified that $Y_{B-L}$ is conserved for all values of $z$ or, equivalently, that the relation $\sum_{i}Y_{\Delta F_i}=\sum_{\alpha}Y_{\Delta\alpha}$ holds. Similarly, the baryon asymmetry vanishes in the absence of flavor effects, \textit{i.e.} when $|\lambda_{e1}|=|\lambda_{\mu 1}|=|\lambda_{\tau 1}|$ and $|\lambda_{e2}|=|\lambda_{\mu 2}|=|\lambda_{\tau 2}|$.

Note that, contrary to the standard leptogenesis scenarios, the SM flavor asymmetries $Y_{\Delta\alpha}$ are constantly generated, since the $CP$-violating processes always occur out-of-equilibrium. They eventually attain their final value as soon as the DM state freezes-in at $z\sim 3-5$. On the other hand, as the heavy leptons decay, their asymmetries $Y_{\Delta F_i}$ decrease and eventually vanish at high $z$. The asymmetry in $F_2$ is many orders of magnitude smaller than the one in $F_1$, due to its smaller couplings $\lambda_{\alpha 2}\ll\lambda_{\alpha 1}$, and is not shown in Figure \ref{fig:Asymmetries}. Once the $F_i$'s have decayed away, conservation of $Y_{B-L}$ implies that also $Y_{B-L_{SM}}\rightarrow 0$. Thus, the model predicts an equal amount of baryon $Y_B$ and SM lepton $Y_{L_{\text{SM}}}$ asymmetries left in the Universe \cite{Gonzalez:2009}.

\begin{figure}[t]
\centering
\includegraphics[width=0.75\linewidth]{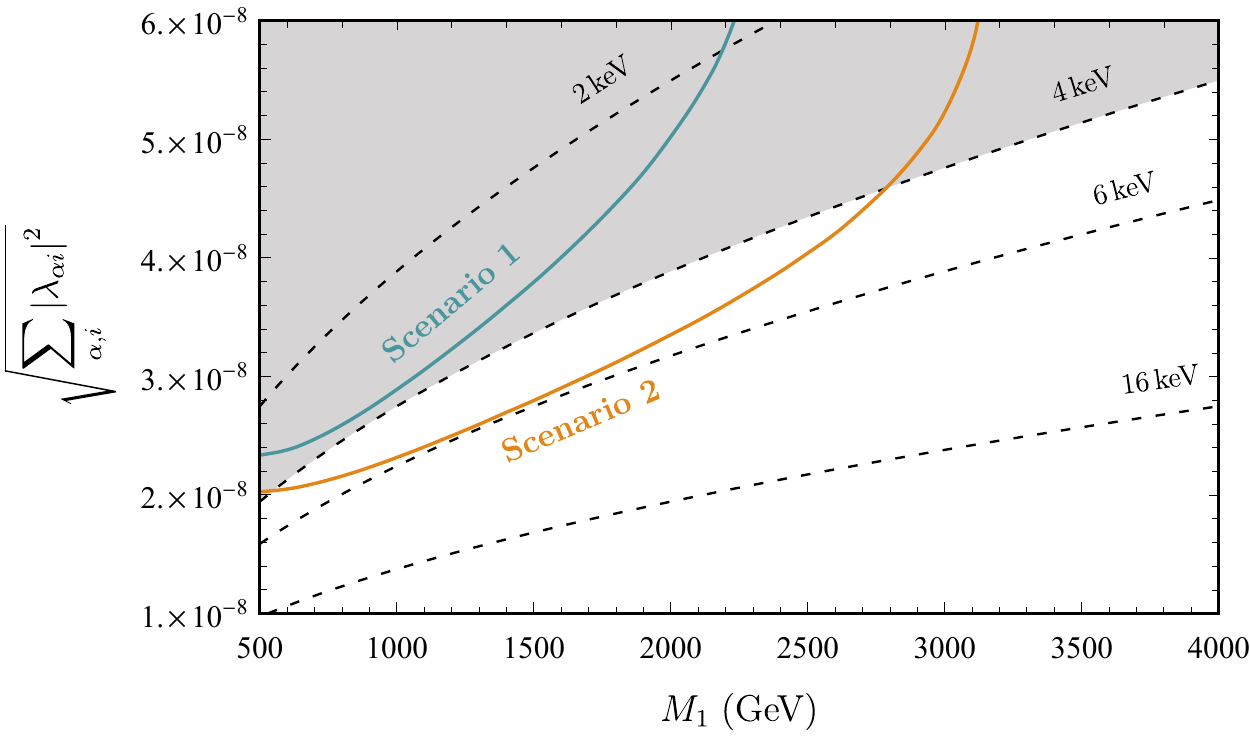}
\caption{Combinations of the feeble couplings and the heavy lepton mass which can generate the observed baryon asymmetry, for the set of parameters depicted in Table \ref{table: scenarios parameters}. The dashed lines depict representative DM masses $m_S=\{2,4,6,16\}\keV$ for which the observed DM abundance of the Universe can be reproduced. Masses below $4\keV$ are excluded from current Lyman-$\alpha$ forest observations (gray shaded area).}
\label{fig:Plot_Baryon_Asymmetry_Scan}
\end{figure}

Our numerical analysis reveals that at least one of the feeble couplings must have a magnitude larger than $\sim 10^{-8}$ in order to account for the observed baryon asymmetry. Our main results are summarized in Figure \ref{fig:Plot_Baryon_Asymmetry_Scan}, where we show the combinations of the feeble couplings and the heavy lepton mass for which a viable baryon asymmetry can be generated, for the two illustrative scenarios depicted in Table \ref{table: scenarios parameters}.

\begin{table}[h]
\centering
\begin{tabular}{l*{5}{c}r}\hline\hline
                  & $\left|\lambda_{\mu 1}\right|$ & $\left|\lambda_{\tau 1}\right|$ & $\left|\lambda_{e2}\right|$ &  $\left|\lambda_{\mu 2}\right|$ &  $\left|\lambda_{\tau 2}\right|$ \\
\hline
Scenario 1        & $10^{-9}$ & $2\times 10^{-9}$ & $9\times 10^{-10}$ & $8\times 10^{-10}$ & $9\times 10^{-10}$   \\
Scenario 2        & $9\times 10^{-9}$ & $9\times 10^{-9}$ & $9\times 10^{-10}$ & $8\times 10^{-10}$ & $9\times 10^{-10}$   \\
\hline\hline
\end{tabular}
\caption{Values of the parameters used for the two scenarios shown in Figure \ref{fig:Plot_Baryon_Asymmetry_Scan}.}
\label{table: scenarios parameters}
\end{table}
 
\noindent
In both cases, the $CP$ asymmetry is resonantly enhanced, \textit{i.e.} the mass of $F_2$ is $M_2=M_1+\Gamma_2/2$ and we consider $p^{12}_{e\mu}=p^{12}_{e\tau}=p^{12}_{\mu\tau}=1$ for the phases of the feeble couplings. The $\lambda_{e1}$ coupling is being treated as a free parameter and lies in the $10^{-8}$ ballpark. 

The dark matter relic density constraint can be satisfied throughout the parameter ranges depicted in the figure, for an appropriate choice of the mass $m_S$. However, in the shaded region, the required dark matter mass turns out to be in conflict with the Lyman-$\alpha$ bound of Eq.~\eqref{eq:Lyman-a bound}. 

We observe that the first scenario (upper, blue line) is excluded from Lyman-$\alpha$ constraints for all values of the heavy lepton mass $M_1$. On the contrary, the second scenario (lower, orange line) is able to account for both the baryon asymmetry and the DM abundance for a wide range of $M_1$, $550\GeV\lesssim M_1\lesssim 2800\GeV$. Note that, interestingly, we find that for any combination of parameter values the relic density constraint is satisfied for DM masses that do not exceed $\sim 6\keV$. This implies that if Lyman-$\alpha$ constraints become stronger in the future, they may be able to fully exclude the model's viable parameter space. 

\subsection{Phenomenological aspects}\label{subsec:pheno}

In the previous paragraphs, we saw that our simple model, described by the Lagrangian of Eq.~\eqref{eq:Lgeneral}, can provide a common explanation for the observed dark matter content and baryon asymmetry of the Universe. As we showed, this can be achieved by assuming highly degenerate heavy fermions $F_i$ with masses $M_{i} \sim 1$ TeV and Yukawa-like couplings of the order of $\lambda_{\alpha 1} \sim 10^{-8}-10^{-7}$ and $\lambda_{\alpha 2} \sim 10^{-10}$ for $F_1$ and $F_2$, respectively. 

Our model is intended to serve mainly as a proof-of-concept for the freeze-in baryogenesis mechanism that we propose. In this spirit, a full-blown analysis of its phenomenological predictions goes well beyond the scope of the present work, especially since alternative constructions assuming, \textit{e.g.}, different quantum numbers for the various particles involved, can lead to wildly different phenomenological signatures. Still, despite its simplicity, this model does exhibit an interesting phenomenology, which we will briefly comment upon. Note that a simpler variant of this model involving a single vectorlike heavy fermion $F$ was already studied, \textit{e.g.}, in \cite{Belanger:2018sti, Brooijmans:2020yij}, mainly from the viewpoint of its predicted dark matter abundance as well as its signatures in new physics searches at the LHC.

First, the heavy fermions can mediate lepton flavor-violating decays of the type $\ell_\beta \rightarrow \ell_\alpha \gamma$. In particular, given the constraint $Br(\mu \rightarrow e \gamma) < 4.2\times 10^{-13}$ \cite{MEG:2016leq} and assuming that (as was the case throughout the analysis presented in the previous sections) $\lambda_{e1} \sim \lambda_{\mu 1}$, the analysis presented in \cite{Brooijmans:2020yij} restricts the product of the Yukawa-like couplings of $F_{1}$ to the first two generation leptons as
\begin{equation}
\frac{\sqrt{\lambda_{e1} \lambda_{\mu 1}}}{M_1} \lesssim 3.6\times 10^{-5} \ {\rm GeV}^{-1}
\end{equation}
which is easily satisfied in the cosmologically relevant part of our parameter space. In deriving this bound, we have neglected the interference between Feynman diagrams involving different species of heavy fermions running in the loop, since the couplings $\lambda_{\alpha 2} \ll \lambda_{\alpha 1}$. Besides, constraints stemming from measurements of the muon lifetime turn out to be subleading \cite{Brooijmans:2020yij}.

On the side of collider searches, let us first point out the fact that, given the feeble nature of the $\lambda_{\alpha i}$ couplings, the mean proper decay length of $F_1$ is of the order of a few centimeters, whereas that of $F_2$ is of the order of several meters. Given these lifetime ranges, both $F_{1}$ and $F_2$ can be looked for in searches for long-lived particles at the LHC. In particular, as shown in \cite{Belanger:2018sti, Brooijmans:2020yij}, searches for displaced leptons accompanied by missing energy can target the production and decay of $F_1$ and, for the values of $\lambda_{\alpha 1}$ that we assume, already exclude $F_1$ masses up to $\sim 400-500$ GeV. $F_2$, on the other hand, survives long enough to escape the detector and can be probed by searches for heavy stable charged particles, with masses smaller than $\sim 500-600$ GeV being already excluded. The corresponding searches at the High-Luminosity Run of the LHC will probe $F_{1,2}$ masses reaching up to $\sim 800$ GeV and $\sim 1.5$ TeV, respectively \cite{Belanger:2018sti}.

In summary, if our model is to simultaneously explain the dark matter abundance and the baryon asymmetry of the Universe while remaining consistent with Lyman-$\alpha$ bounds, then it predicts interesting signatures for LLP searches at the LHC. In conjunction with the fact that, as we showed, explaining the matter-antimatter asymmetry of the Universe tends to favor $F_{1,2}$ masses lying in the range of a few TeV, we can hope that a substantial part of the cosmologically favored part of the parameter space will be probed by the LHC within the next few years.
\section{Conclusions and Outlook}\label{sec:conclusions}

In this paper, we discussed a mechanism for the simultaneous generation of the dark matter density and the baryon asymmetry of the Universe. To this goal, we relied on the out-of-equilibrium decays of heavy bath particles into a (feebly coupled) dark matter state along with Standard Model charged fermions, which leads to dark matter production via the freeze-in mechanism. If, moreover, $CP$ is violated by these same decay processes, a viable matter-antimatter asymmetry can also be generated either directly (if the decays also lead to baryon/lepton number violation) or through the interplay of the generated $CP$ asymmetry with the SM electroweak sphalerons. 

As a proof-of-concept, we employed a simple model in which the role of the heavy bath particles was played by two $SU(3)_c\times SU(2)_{\LH}$-singlet vectorlike fermions with a non-zero hypercharge, and dark matter was identified with a gauge singlet real scalar field. We showed that, indeed, such a simple construction can lead to both dark matter production and successful baryogenesis and we briefly discussed the phenomenological prospects of such a construction, particularly in relation to searches for long-lived particles at the LHC.

Our proposal draws inspiration from the idea presented in \cite{Shuve_2020}: First, the dark matter content and the baryon asymmetry of the Universe are generated simultaneously through the freeze-in mechanism. Second, at least within the framework of our concrete incarnation of this proposal, in both cases the electroweak sphalerons are used in order to convert a $CP$ asymmetry into a baryonic one. In our case, however, instead of considering oscillations -- as in ARS leptogenesis -- we rather relied on decays in order to generate this initial $CP$ asymmetry -- a process which is somewhat reminiscent of GUT baryogenesis. Moreover, we presented a general thermodynamical treatment that can also be applied in cases in which baryon/lepton number is violated already at the decay level.

There are several topics which we did not expand upon in this paper, and which would merit much more detailed investigation. First, more elaborate models can certainly be developed in this context of freeze-in baryogenesis, potentially establishing connections with theoretically well-motivated UV completions of the Standard Model. As an example, we could mention the fact that we assumed rather \textit{ad hoc} values for the couplings of the heavy bath particles to dark matter and to the SM fermions. More realistic flavor structures can be envisaged in quite a straightforward manner, whereas nothing forbids dark matter itself to be asymmetric, along the lines described in \cite{Hall:2010jx}. Second, albeit related to the previous point, our discussion of more phenomenological aspects has been fairly elementary. Concrete, well-motivated constructions can have important implications for flavor physics (\textit{e.g.} flavor-violating decays of or contributions to electric/magnetic moments of SM fermions), collider physics (\textit{e.g.} new resonances and/or final states in LHC searches, potentially involving long-lived particles \cite{Alimena:2019zri}), or cosmological measurements (\textit{e.g.} direct detection experiments or cosmic microwave background observations). Such considerations, which are typically fairly model dependent, are left for future work.

\acknowledgments 
The authors acknowledge fruitful exchanges with Genevi\`eve B\'elanger, Marcos A. G. Garcia, Kalliopi Petraki and Alexander Pukhov. We are particularly grateful toward Sacha Davidson for enlightening discussions and comments on an earlier version of this manuscript and toward Dimitrios Karamitros for useful comments on our findings. This research is co-financed by Greece and the European Union (European Social Fund - ESF) through the  Operational Programme \textquote{Human Resources Development, Education and Lifelong Learning} in the context of the project \textquote{Strengthening Human Resources Research Potential via Doctorate Research - 2nd Cycle} (MIS-5000432), implemented by the State Scholarships Foundation (IKY). This research work was supported by the Hellenic Foundation for Research and Innovation (H.F.R.I.) under the ``First Call for H.F.R.I. Research Projects to support Faculty members and Researchers and the procurement of high-cost research equipment grant'' (Project Number: 824).
   
\begin{appendix}

\section{Spectator processes}\label{app:Spectator Processes}

During the baryon and/or lepton number violation era, there are various processes which can modify the number densities of the particles species. They are called "spectator processes" because they affect the baryon/lepton number indirectly by distributing the asymmetry among all SM fermions and Higgs fields. These processes include the gauge interactions, the Yukawa interactions and the electroweak and QCD non-perturbative sphaleron transitions. If these processes are in thermal equilibrium with the cosmic plasma, then their effect is to impose certain relations among the chemical potentials of the various particle species. Recall that the asymmetry in the particle and antiparticle equilibrium number densities of species $k$, denoted as $n_{\Delta k}\equiv n_{k}-n_{\bar{k}}$, is given in the ultra-relativistic limit $m_{k}\ll T$ and $\mu_{k}\ll T$ by

\begin{align}\label{eq:number_density_asymmetry}
n_{\Delta k}\,\equiv\,n_{k}-n_{\bar{k}}\,=\,\left\{\,
\begin{array}{@{}ll@{}}
    g_{k}\mu_{k}\frac{T^2}{3},\qquad &k:\text{boson}
    \\\\
    g_{k}\mu_{k}\frac{T^2}{6},\qquad &k:\text{fermion}
	\end{array}
\right.
\end{align}
\\
\noindent
where $\mu_{k}$ is the chemical potential and $g_{k}$ are the independent degrees of freedom of species $k$. The total baryon $n_B$ and lepton $n_L$ asymmetries can be written as

\begin{equation}
n_B\,=\,\sum_{k}\,B_{k}\,n_{\Delta k},\qquad n_L\,=\,\sum_{k}\,L_{k}\,n_{\Delta k}
\end{equation}

\noindent
where $B_{k}$ and $\,L_{k}$ are the baryon and lepton numbers of species $k$, respectively. 

In our case, the lepton asymmetry is mainly generated during $T_{EW}\leq T\ll 10^{8}\,\GeV$ when all spectator processes are active, including the sphaleron processes. These violate the $n_{B+L}\equiv n_B+n_L$ asymmetry but conserve the orthogonal combination $n_{B-L}\equiv n_B-n_L$. A rough estimation is that sphalerons will erase the $n_{B+L}$ asymmetry, resulting in

\begin{subequations}
\begin{alignat}{2}
n_B\,=\,\frac{n_{B+L}}{2}\,+\,\frac{n_{B-L}}{2}\,&\overset{\text{sphalerons}}{\longrightarrow}\,n_B\,=\,\frac{n_{B-L}}{2}
\\
n_L\,=\,\frac{n_{B+L}}{2}\,-\,\frac{n_{B-L}}{2}\,&\overset{\text{sphalerons}}{\longrightarrow}\,n_L\,=\,-n_B\,=\,-\frac{n_{B-L}}{2}
\end{alignat}
\end{subequations}
\\
\noindent
A more careful treatment shows that, in spite of the fact that sphalerons violate $n_{B+L}$, thermodynamic equilibrium requires a non-zero value, \textit{i.e.} $n_{B+L}\neq 0$. To quantify the asymmetry distribution, one can always relate the $n_{B-L}$ number density asymmetry with $n_{B}$ and $n_{L}$ as follows:

\begin{equation}
n_{B-L}\,\equiv\,n_{B}\,-\,n_{L}\,=\,c\,n_{B-L}\,-\,\left(c-1\right)\,n_{B-L}\,\Rightarrow\,\left\{\,\begin{array}{@{}ll@{}}
    n_{B}\,=\,c\,n_{B-L}\\\\
    n_{L}\,=\,\left(c-1\right)\,n_{B-L}
	\end{array}
\right.
\end{equation}

\noindent
where $c\lesssim 1$ is the constant of proportionality. To evaluate it, we assign chemical potentials to the particle spectrum of the model. This consists of all the SM species with three families and one Higgs doublet, together with the two BSM vectorlike fermions $F_i$.

\subsubsection{SM}

Let us first consider only the SM spectrum and assign the following non-vanishing chemical potentials \cite{Harvey:1990qw,Buchmuller:2005eh}

\begin{subequations}\label{eq:SM_chemical_potentials}
\begin{alignat}{5}
\mu_{q_{\alpha}}\qquad&\left(\text{LH quark doublets}\right)
\\
\mu_{u_{\alpha}}\qquad&\left(\text{RH up-quark singlets}\right)
\\
\mu_{d_{\alpha}}\qquad&\left(\text{RH down-quark singlets}\right)
\\
\mu_{{\ell}_{\alpha}}\qquad&\left(\text{LH lepton doublets}\right)
\\
\mu_{e_{\alpha}}\qquad&\left(\text{RH charged lepton singlets}\right)
\\
\mu_H\qquad&\left(\text{Higgs doublet,}\,H=\left(h^+,h^0\right)^T\right)
\end{alignat}
\end{subequations}

\noindent
where $\alpha=\{1,2,3\}$ is a SM flavor index.\footnote{Note that in this Appendix we are introducing a slightly different notation with respect to the rest of the paper, denoting the SM right-handed leptons by $\{e_1,e_2,e_3\}$ instead of $\{e_e,e_{\mu},e_{\tau}\}$.} The corresponding antiparticle states have opposite chemical potentials, \textit{e.g.} $\mu_{\tilde{H}}=-\mu_{H}$, where $\tilde{H}=\left(h^{0*},-h^{-}\right)^T$. Note that the electrically chargeless gauge bosons have vanishing chemical potentials, as they carry no conserved quantum number, while the chemical potential of the $W^{\pm}$ gauge bosons vanishes, because the third - component of the weak isospin is zero at $T>T_{EW}$ \cite{Harvey:1990qw}. Hence, the number density asymmetries of the various components can be written, according to Eq.~\eqref{eq:number_density_asymmetry}, as

\begin{subequations}\label{eq:SM_asymmetric_densities}
\begin{alignat}{6}
n_{\Delta q_{\alpha}}\,&\equiv\,n_{q_{\alpha}}\,-\,n_{\bar{q}_{\alpha}}\,=\,6\,\mu_{q_{\alpha}}\,\frac{T^2}{6}
\\
n_{\Delta u_{\alpha}}\,&\equiv\,n_{u_{\alpha}}\,-\,n_{\bar{u}_{\alpha}}\,=\,3\,\mu_{u_{\alpha}}\,\frac{T^2}{6}
\\
n_{\Delta d_{\alpha}}\,&\equiv\,n_{d_{\alpha}}\,-\,n_{\bar{d}_{\alpha}}\,=\,3\,\mu_{d_{\alpha}}\,\frac{T^2}{6}
\\
n_{\Delta {\ell}_{\alpha}}\,&\equiv\,n_{{\ell}_{\alpha}}\,-\,n_{\bar{\ell}_{\alpha}}\,=\,2\,\mu_{{\ell}_{\alpha}}\,\frac{T^2}{6}
\\
n_{\Delta e_{\alpha}}\,&\equiv\,n_{e_{\alpha}}\,-\,n_{\bar{e}_{\alpha}}\,=\,\mu_{e_{\alpha}}\,\frac{T^2}{6}
\\
n_{\Delta H}\,&\equiv\,n_H\,-\,n_{\tilde{H}}\,=\,2\,\mu_H\,\frac{T^2}{3}\,=\,4\,\mu_H\,\frac{T^2}{6}
\end{alignat}
\end{subequations}

\noindent
The baryon and lepton number density asymmetries for each flavor $\left(n_{B_{\alpha}},\,n_{L_{\alpha}}\right)$, as well as the corresponding total ones $\left(n_B,\,n_L\right)$, can be written in terms of the chemical potentials in Eq.~\eqref{eq:SM_chemical_potentials} as follows:

\begin{subequations}
\begin{alignat}{4}
n_{B_{\alpha}}\,&\equiv\,\sum_{k}\,B_{k_{\alpha}}\,n_{\Delta k_{\alpha}}\,=\,\frac{1}{3}\left(n_{\Delta q_{\alpha}}\,+\,n_{\Delta u_{\alpha}}\,+\,n_{\Delta d_{\alpha}}\right)\,=\,\left(2\mu_{q_{\alpha}}\,+\,\mu_{u_{\alpha}}\,+\,\mu_{d_{\alpha}}\right)\,\frac{T^2}{6}
\\
n_B\,&\equiv\,\sum_{\alpha=1}^3\,n_{B_{\alpha}}\,=\,\sum_{\alpha=1}^3\,\left(2\mu_{q_{\alpha}}\,+\,\mu_{u_{\alpha}}\,+\,\mu_{d_{\alpha}}\right)\,\frac{T^2}{6}
\\
n_{L_{\alpha}}\,&\equiv\,\sum_{k}\,L_{k_{\alpha}}\,n_{\Delta k_{\alpha}}\,=\,n_{\Delta\ell_{\alpha}}\,+\,n_{\Delta e_{\alpha}}\,=\,\left(2\mu_{\ell_{\alpha}}\,+\,\mu_{e_{\alpha}}\right)\,\frac{T^2}{6}
\\
n_{L}\,&\equiv\,\sum_{\alpha=1}^3\,n_{L_{\alpha}}\,=\,\sum_{\alpha=1}^3\left(2\mu_{\ell_{\alpha}}\,+\,\mu_{e_{\alpha}}\right)\,\frac{T^2}{6}
\end{alignat}
\end{subequations}

\noindent
These 16 SM chemical potentials are not totally independent; they are related through the processes that attain chemical equilibrium during the era of the asymmetry generation. At $T\ll 10^8$ these are

\begin{itemize}
\item $SU(2)_L$ sphaleron transitions which induce an effective 12-fermion effective operator $\mathcal{O}_{(B-L)_{\text{SM}}}\,=\,\prod_{i=1}^{N_f}\left(q_iq_iq_i\ell_i\right)$, which implies

\begin{equation}\label{eq:SU(2)_sphaleron_constraint}
\sum_{\alpha=1}^3\left(3\mu_{q_{\alpha}}\,+\,\mu_{\ell_{\alpha}}\right)\,=\,0
\end{equation}

\item $SU(3)_c$ sphaleron transitions which give rise to

\begin{equation}\label{eq:SU(3)_sphaleron_constraint}
\sum_{\alpha=1}^3\left(2\mu_{q_{\alpha}}\,-\,\mu_{u_{\alpha}}\,-\,\mu_{d_{\alpha}}\right)\,=\,0
\end{equation}

\item Yukawa interactions which imply

\begin{subequations}\label{eq:Yukawa_constraints}
\begin{alignat}{3}
\mu_{q_{\alpha}}\,-\,\mu_H\,-\,\mu_{d_{\alpha}}\,=\,0,\qquad&\left(\,q_{\alpha}\,\tilde{H}\,\bar{d}_{\alpha}\,+\,\text{h.c.}\,\right)\label{eq:Yukawa_1}
\\
\mu_{q_{\alpha}}\,+\,\mu_H\,-\,\mu_{u_{\alpha}}\,=\,0,\qquad&\left(\,q_{\alpha}\,H\,\bar{u}_{\alpha}\,+\,\text{h.c.}\,\right)\label{eq:Yukawa_2}
\\
\mu_{{\ell}_{\alpha}}\,-\,\mu_H\,-\,\mu_{e_{\alpha}}\,=\,0,\qquad&\left(\,{\ell}_{\alpha}\,\tilde{H}\,\bar{e}_{\alpha}\,+\,\text{h.c.}\,\right)\label{eq:Yukawa_3}
\end{alignat}
\end{subequations}

\end{itemize}

Out of the 11 constraints of Eqs.~\eqref{eq:SU(2)_sphaleron_constraint} - \eqref{eq:Yukawa_constraints}, only ten of them are linearly independent, as the QCD sphaelron-induced relation \eqref{eq:SU(3)_sphaleron_constraint} can be obtained by adding Eqs.~\eqref{eq:Yukawa_1} and \eqref{eq:Yukawa_2}. An additional independent constraint can be obtained from the hypercharge $\left(\Upsilon=Q-t_3\right)$ neutrality condition, which implies

\begin{itemize}

\item Hypercharge constraint

\begin{align}\label{eq:hypercharge_constraint}
n_{\Upsilon}\,&\equiv\,\sum_{k}\,\Upsilon_{k}\,n_{\Delta k}\,=\,0\nonumber
\\
&\Rightarrow\,\sum_{\alpha=1}^3\Big(\frac{1}{6}n_{\Delta q_{\alpha}}\,+\,\frac{2}{3}n_{\Delta u_{\alpha}}\,-\,\frac{1}{3}n_{\Delta d_{\alpha}}\,-\,\frac{1}{2}\,n_{\Delta\ell_{\alpha}}\,-\,n_{\Delta e_{\alpha}}\Big)\,+\,\frac{1}{2}n_{\Delta H}\,=\,0\nonumber
\\
&\Rightarrow\,\sum_{\alpha=1}^3\left(\mu_{q_{\alpha}}\,+\,2\mu_{u_{\alpha}}\,-\,\mu_{d_{\alpha}}\,-\,\mu_{\ell_{\alpha}}\,-\,\mu_{e_{\alpha}}\right)\,+\,2\mu_H\,=\,0
\end{align}

\end{itemize}

\noindent
where $\Upsilon_{k}$ is the hypercharge of species $k$. Two more constraints are imposed by the equality of the baryon flavor asymmetries \cite{Davidson:2008bu}.

\begin{itemize}

\item Baryon flavor asymmetry equality

\begin{align}\label{eq:baryon_flavor_constraints}
& n_{B_1}\,=\,n_{B_2}\,=\,n_{B_3}\nonumber
\\\nonumber
\\
\Rightarrow\,& 2\,\mu_{q_3}\,+\,\mu_{u_3}\,+\,\mu_{d_3}\,=\,2\,\mu_{q_2}\,+\,\mu_{u_2}\,+\,\mu_{d_2}\,=\,2\,\mu_{q_1}\,+\,\mu_{u_1}\,+\,\mu_{d_1}
\end{align}

\end{itemize}

\noindent
Hence, there are in total 13 independent constraints (Eqs.~\eqref{eq:SU(2)_sphaleron_constraint}, \eqref{eq:Yukawa_constraints}, \eqref{eq:hypercharge_constraint} and \eqref{eq:baryon_flavor_constraints}) and, therefore, one can express the 16 SM chemical potentials \eqref{eq:SM_chemical_potentials} in terms of three chemical potentials, that we chose to be $\{\mu_{\ell_{\alpha}}\}$. Solving the system of equations, one obtains

\begin{subequations}\label{eq:SM_chemical_potentials_constraints}
\begin{alignat}{5}
\mu_{q_{\alpha}}\,&=\,-\frac{1}{9}\left(\mu_{\ell_1}\,+\,\mu_{\ell_2}\,+\,\mu_{\ell_3}\right)
\\
\mu_{u_{\alpha}}\,&=\,\frac{5}{63}\left(\mu_{\ell_1}\,+\,\mu_{\ell_2}\,+\,\mu_{\ell_3}\right)
\\
\mu_{d_{\alpha}}\,&=\,-\frac{19}{63}\left(\mu_{\ell_1}\,+\,\mu_{\ell_2}\,+\,\mu_{\ell_3}\right)
\\
\mu_{e_1}\,&=\,\frac{1}{21}\left(17\mu_{\ell_1}\,-\,4\mu_{\ell_2}\,-\,4\mu_{\ell_3}\right)
\\\mu_{e_2}\,&=\,\frac{1}{21}\left(-4\mu_{\ell_1}\,+\,17\mu_{\ell_2}\,-\,4\mu_{\ell_3}\right)
\\
\mu_{e_3}\,&=\,\frac{1}{21}\left(-4\mu_{\ell_1}\,-\,4\mu_{\ell_2}\,+\,17\mu_{\ell_3}\right)
\\
\mu_{H}\,&=\,\frac{4}{21}\left(\mu_{\ell_1}\,+\,\mu_{\ell_2}\,+\,\mu_{\ell_3}\right)
\end{alignat}
\end{subequations}
\\
\noindent
Now the baryon and lepton number density asymmetries can be written in terms of the chemical potentials $\mu_{\ell_{\alpha}}$ as

\begin{subequations}
\begin{alignat}{4}
n_{B_{\alpha}}\,&=\,\left(2\mu_{q_{\alpha}}\,+\,\mu_{u_{\alpha}}\,+\,\mu_{d_{\alpha}}\right)\,\frac{T^2}{6}\,=\,-\frac{4}{9}\left(\mu_{\ell_1}\,+\,\mu_{\ell_2}\,+\,\mu_{\ell_3}\right)\,\frac{T^2}{6}
\\
n_B\,&=\,3\,n_{B_{\alpha}}\,=\,-\frac{4}{3}\left(\mu_{\ell_1}\,+\,\mu_{\ell_2}\,+\,\mu_{\ell_3}\right)\,\frac{T^2}{6}
\\
n_{L_{\alpha}}\,&=\,\left(2\mu_{\ell_{\alpha}}\,+\,\mu_{e_{\alpha}}\right)\,\frac{T^2}{6}
\\
n_{L}\,&=\,\sum_{\alpha=1}^3\left(2\mu_{\ell_{\alpha}}\,+\,\mu_{e_{\alpha}}\right)\,\frac{T^2}{6}\,=\,\frac{17}{7}\left(\mu_{\ell_1}\,+\,\mu_{\ell_2}\,+\,\mu_{\ell_3}\right)\,\frac{T^2}{6}
\end{alignat}
\end{subequations}

\noindent
The flavor asymmetry combination $n_{\Delta_{\alpha}}\equiv n_{B_{\alpha}}-n_{L_{\alpha}}=n_{B}/3-n_{L_{\alpha}}$ can be written as

\begin{flalign*}
n_{\Delta_{\alpha}}\,&=\,\left(2\mu_{q_{\alpha}}\,+\,\mu_{u_{\alpha}}\,+\,\mu_{d_{\alpha}}\,-\,2\mu_{\ell_{\alpha}}\,-\,\mu_{e_{\alpha}}\right)\,\frac{T^2}{6} &
\\\nonumber
\\
& \Rightarrow\,
\begin{pmatrix}
n_{\Delta_1} \\ 
n_{\Delta_2} \\ 
n_{\Delta_3}
\end{pmatrix}
\,=\,
-\frac{1}{63}\,
\left[\begin{array}{ccc}
205 & 16 & 16 \\
16 & 205 & 16 \\
16 & 16 & 205
\end{array}\right]\,
\begin{pmatrix}
\mu_{\ell_1} \\ 
\mu_{\ell_2} \\ 
\mu_{\ell_3}
\end{pmatrix}\,\frac{T^2}{6} &
\end{flalign*}
\\
\noindent
Inverting the expression above, one finds

\begin{equation}
\begin{pmatrix}
\mu_{\ell_1} \\ 
\mu_{\ell_2} \\ 
\mu_{\ell_3}
\end{pmatrix}
\,=\,
-\frac{1}{711}\,
\left[\begin{array}{ccc}
221 & -16 & -16 \\
-16 & 221 & -16 \\
-16 & -16 & 221
\end{array}\right]\,
\begin{pmatrix}
n_{\Delta_1} \\ 
n_{\Delta_2} \\ 
n_{\Delta_3}
\end{pmatrix}\,\frac{6}{T^2}
\end{equation}
\\
\noindent
Using Eq.~\eqref{eq:SM_asymmetric_densities}, we can express the chemical potentials $\mu_{\ell_{\alpha}}$ in terms of the flavor asymmetries $n_{\Delta\ell_{\alpha}}$ and relate them to $n_{\Delta\alpha}$ as \footnote{Note that our result agrees with reference \cite{Nardi_2006}, while the authors of reference \cite{Gonzalez:2009} define $n_{\Delta\ell_{\alpha}}$ per gauge degree of freedom and, therefore, their result is smaller by a factor of 2.}

\begin{equation}
\begin{pmatrix}
n_{\Delta\ell_1} \\ 
n_{\Delta\ell_2} \\ 
n_{\Delta\ell_3}
\end{pmatrix}
\,=\,
-\frac{2}{711}\,
\left[\begin{array}{ccc}
221 & -16 & -16 \\
-16 & 221 & -16 \\
-16 & -16 & 221
\end{array}\right]\,
\begin{pmatrix}
n_{\Delta_1} \\ 
n_{\Delta_2} \\ 
n_{\Delta_3}
\end{pmatrix}
\end{equation}
\\
\noindent
The total $B-L$ asymmetry is

\begin{equation}
n_{B-L}\,\equiv\,n_B\,-\,n_L\,=\,-\frac{79}{21}\left(\mu_{\ell_1}\,+\,\mu_{\ell_2}\,+\,\mu_{\ell_3}\right)\,\frac{T^2}{6}
\end{equation}

\noindent
Hence, in the SM and for $T_{sph}\ll T\ll 10^{8}\,\GeV$, the total baryon $n_B$ and the $n_L$ asymmetries are related to $n_{B-L}$ by

\begin{align}
n_B\,=\,\frac{28}{79}\,n_{B-L}\,=\,\frac{28}{79}\,\sum_{\alpha=1}^3n_{\Delta\alpha}
\\
n_L\,=\,-\frac{51}{79}\,n_{B-L}\,=\,-\frac{51}{79}\,\sum_{\alpha=1}^3n_{\Delta\alpha}
\end{align}
\\

\subsubsection{Our model}

We extend the SM to include also two heavy vectorlike leptons $F_1$ and $F_2$ with chemical potentials $\mu_{F_1}$ and $\mu_{F_2}$, respectively, and asymmetry

\begin{equation}
n_{\Delta F_i}\,=\,\mu_{F_i}\,\frac{T^2}{6}
\end{equation}

\noindent
Since the BSM fermions are gauged under $U(1)_{\Upsilon}$, with $\Upsilon_{F_i}=\Upsilon_{e}=-1$, the hypercharge constraint \eqref{eq:hypercharge_constraint} is modified to

\begin{align}\label{eq:hypercharge_constraint_modified}
n_{\Upsilon}\,&\equiv\,\sum_{k}\,\Upsilon_{k}\,n_{\Delta k}\,=\,0\nonumber
\\
&\Rightarrow\,\sum_{\alpha=1}^3\Big(\frac{1}{6}n_{\Delta q_{\alpha}}\,+\,\frac{2}{3}n_{\Delta u_{\alpha}}\,-\,\frac{1}{3}n_{\Delta d_{\alpha}}\,-\,\frac{1}{2}\,n_{\Delta\ell_{\alpha}}\,-\,n_{\Delta e_{\alpha}}\Big)\,+\,\frac{1}{2}n_{\Delta H}\,-\,\sum_{i=1}^2\,n_{\Delta F_i}\,=\,0\nonumber
\\
&\Rightarrow\,\sum_{\alpha=1}^3\left(\mu_{q_{\alpha}}\,+\,2\mu_{u_{\alpha}}\,-\,\mu_{d_{\alpha}}\,-\,\mu_{\ell_{\alpha}}\,-\,\mu_{e_{\alpha}}\right)\,+\,2\mu_H\,-\,\left(\mu_{F_1}\,+\,\mu_{F_2}\right)\,=\,0
\end{align} 

\noindent
The theory conserves the total $n_{B-L}\equiv n_B-n_L= n_B-n_{L_{SM}}-n_{L_F}$ asymmetry, where

\begin{subequations}
\begin{alignat}{4}
n_{B_{\alpha}}\,&\equiv\,\sum_{k}\,B_{k_{\alpha}}\,n_{\Delta k_{\alpha}}\,=\,\frac{1}{3}\left(n_{\Delta q_{\alpha}}\,+\,n_{\Delta u_{\alpha}}\,+\,n_{\Delta d_{\alpha}}\right)\,=\,\left(2\mu_{q_{\alpha}}\,+\,\mu_{u_{\alpha}}\,+\,\mu_{d_{\alpha}}\right)\,\frac{T^2}{6}
\\
n_B\,&\equiv\,\sum_{\alpha=1}^3\,n_{B_{\alpha}}\,=\,\sum_{\alpha=1}^3\,\left(2\mu_{q_{\alpha}}\,+\,\mu_{u_{\alpha}}\,+\,\mu_{d_{\alpha}}\right)\,\frac{T^2}{6}\,=\,3\,\left(2\mu_{q_{\alpha}}\,+\,\mu_{u_{\alpha}}\,+\,\mu_{d_{\alpha}}\right)\,\frac{T^2}{6}
\\
n_{L_{\alpha}}\,&\equiv\,\sum_{k:\,SM}\,L_{k_{\alpha}}\,n_{\Delta k_{\alpha}}\,=\,n_{\Delta\ell_{\alpha}}\,+\,n_{\Delta e_{\alpha}}\,=\,\left(2\mu_{\ell_{\alpha}}\,+\,\mu_{e_{\alpha}}\right)\,\frac{T^2}{6}
\\
n_{L_{SM}}\,&\equiv\,\sum_{\alpha=1}^3\,n_{L_{\alpha}}\,=\,\sum_{\alpha=1}^3\left(2\mu_{\ell_{\alpha}}\,+\,\mu_{e_{\alpha}}\right)\,\frac{T^2}{6}
\\
n_{L_F}\,&\equiv\,\sum_{i=1}^2\,n_{\Delta F_i}\,=\,\sum_{i=1}^2\,\mu_{F_i}\,\frac{T^2}{6}
\\
n_L\,&\equiv\,n_{L_{SM}}\,+\,n_{L_F}\,=\,\left(\sum_{\alpha=1}^3\left(2\mu_{\ell_{\alpha}}\,+\,\mu_{e_{\alpha}}\right)\,+\,\sum_{i=1}^2\,\mu_{F_i}\right)\,\frac{T^2}{6}
\end{alignat}
\end{subequations}

\noindent
If we also assume that the early Universe is totally symmetric, $n_{B-L_{SM}}|_0=0$ and $n_F|_0=0$, then we obtain an additional constraint to $\mu_{F_1}+\mu_{F_2}$, that is

\begin{align}\label{eq:B-L_conservation_constraint}
& n_B\,-\,n_{L_{SM}}\,-\,n_{L_F}\,=\,\sum_{\alpha=1}^3\,n_{\Delta\alpha}\,-\,\sum_{i=1}^2\,n_{\Delta F_i}\,=\,0\nonumber
\\
\Rightarrow &\sum_{\alpha=1}^3\left(2\mu_{q_{\alpha}}\,+\,\mu_{u_{\alpha}}\,+\,\mu_{d_{\alpha}}\,-\,2\mu_{\ell_{\alpha}}\,-\,\mu_{e_{\alpha}}\right)\,-\,\left(\mu_{F_1}\,+\,\mu_{F_2}\right)\,=\,0
\end{align}

\noindent
Replacing this in Eq.~\eqref{eq:hypercharge_constraint_modified}, we obtain the modified hypercharge constraint expressed in terms of the SM chemical potentials.

\begin{itemize}

\item Modified hypercharge constraint

\begin{equation}\label{eq:hypercharge_constraint_modified_2}
\sum_{\alpha=1}^3\left(-\,\mu_{q_{\alpha}}\,+\,\mu_{u_{\alpha}}\,-\,2\mu_{d_{\alpha}}\,+\,\mu_{\ell_{\alpha}}\right)\,+\,2\mu_H\,=\,0
\end{equation}

\end{itemize}

\noindent
Now the 16 SM chemical potentials are constrained under the $SU(2)$ sphaleron \eqref{eq:SU(2)_sphaleron_constraint} and Yukawa \eqref{eq:Yukawa_constraints} processes (10 constraints), as well as under the baryon flavor asymmetry equality \eqref{eq:baryon_flavor_constraints} and the modified hypercharge \eqref{eq:hypercharge_constraint_modified_2} conditions (3 constraints). Solving the system of equations, we may express all chemical potentials in terms of $\mu_{e_{\alpha}}$

\begin{subequations}\label{eq:BSM_chemical_potentials_constraints}
\begin{alignat}{5}
\mu_{q_{\alpha}}\,&=\,-\frac{11}{144}\left(\mu_{e_1}\,+\,\mu_{e_2}\,+\,\mu_{e_3}\right)
\\
\mu_{u_{\alpha}}\,&=\,-\frac{13}{72}\left(\mu_{e_1}\,+\,\mu_{e_2}\,+\,\mu_{e_3}\right)
\\
\mu_{d_{\alpha}}\,&=\,\frac{1}{36}\left(\mu_{e_1}\,+\,\mu_{e_2}\,+\,\mu_{e_3}\right)
\\
\mu_{\ell_1}\,&=\,\frac{1}{48}\left(43\mu_{e_1}-5\mu_{e_2}-5\mu_{e_3}\right)
\\\mu_{\ell_2}\,&=\,\frac{1}{48}\left(-5\mu_{e_1}+43\mu_{e_2}-5\mu_{e_3}\right)
\\
\mu_{\ell_3}\,&=\,\frac{1}{48}\left(-5\mu_{e_1}-5\mu_{e_2}+43\mu_{e_3}\right)
\\
\mu_{H}\,&=\,-\frac{5}{48}\left(\mu_{e_1}\,+\,\mu_{e_2}\,+\,\mu_{e_3}\right)
\end{alignat}
\end{subequations}
\\
\noindent
Now the baryon and lepton number density asymmetries can be written in terms of the chemical potentials $\mu_{e_{\alpha}}$ as

\begin{subequations}
\begin{alignat}{4}
n_{B_{\alpha}}\,&=\,\left(2\mu_{q_{\alpha}}\,+\,\mu_{u_{\alpha}}\,+\,\mu_{d_{\alpha}}\right)\,\frac{T^2}{6}\,=\,-\frac{11}{36}\left(\mu_{e_1}\,+\,\mu_{e_2}\,+\,\mu_{e_3}\right)\,\frac{T^2}{6}
\\
n_B\,&=\,3\,n_{B_{\alpha}}\,=\,-\frac{11}{12}\left(\mu_{e_1}\,+\,\mu_{e_2}\,+\,\mu_{e_3}\right)\,\frac{T^2}{6}
\\
n_{L_{\alpha}}\,&=\,\left(2\mu_{\ell_{\alpha}}\,+\,\mu_{e_{\alpha}}\right)\,\frac{T^2}{6}
\\
n_{L_{SM}}\,&=\,\sum_{\alpha=1}^3\left(2\mu_{\ell_{\alpha}}\,+\,\mu_{e_{\alpha}}\right)\,\frac{T^2}{6}\,=\,\frac{19}{8}\left(\mu_{e_1}\,+\,\mu_{e_2}\,+\,\mu_{e_3}\right)\,\frac{T^2}{6}
\\
n_{L_F}\,&=\,\left(\mu_{F_1}\,+\,\mu_{F_2}\right)\,\frac{T^2}{6}\,=\,\sum_{\alpha=1}^3\left(2\mu_{q_{\alpha}}\,+\,\mu_{u_{\alpha}}\,+\,\mu_{d_{\alpha}}\,-\,2\mu_{\ell_{\alpha}}\,-\,\mu_{e_{\alpha}}\right)\,\frac{T^2}{6}\nonumber
\\
& =\,-\frac{79}{24}\left(\mu_{e_1}\,+\,\mu_{e_2}\,+\,\mu_{e_3}\right)\,\frac{T^2}{6}\,=\,n_B\,-\,n_{L_{SM}}
\\
n_L\,&\equiv\,n_{L_{SM}}\,+\,n_{L_F}\,=\,-\frac{11}{12}\left(\mu_{e_1}\,+\,\mu_{e_2}\,+\,\mu_{e_3}\right)\,\frac{T^2}{6}\,=\,n_B
\end{alignat}
\end{subequations}
\\
\noindent
The flavor asymmetry combination $n_{\Delta_{\alpha}}\equiv n_{B_{\alpha}}-n_{L_{\alpha}}=n_{B}/3-n_{L_{\alpha}}$ can be written as

\begin{flalign*}
n_{\Delta_{\alpha}}\,&=\,\left(2\mu_{q_{\alpha}}\,+\,\mu_{u_{\alpha}}\,+\,\mu_{d_{\alpha}}\,-\,2\mu_{\ell_{\alpha}}\,-\,\mu_{e_{\alpha}}\right)\,\frac{T^2}{6} &
\\\nonumber
\\
& \Rightarrow\,
\begin{pmatrix}
n_{\Delta_1} \\ 
n_{\Delta_2} \\ 
n_{\Delta_3}
\end{pmatrix}
\,=\,
-\frac{1}{72}\,
\left[\begin{array}{ccc}
223 & 7 & 7 \\
7 & 223 & 7 \\
7 & 7 & 223
\end{array}\right]\,
\begin{pmatrix}
\mu_{e_1} \\ 
\mu_{e_2} \\ 
\mu_{e_3}
\end{pmatrix}\,\frac{T^2}{6} &
\end{flalign*}
\\
\noindent
Inverting the expression above, one finds

\begin{equation}
\begin{pmatrix}
\mu_{e_1} \\ 
\mu_{e_2} \\ 
\mu_{e_3}
\end{pmatrix}
\,=\,
-\frac{1}{711}\,
\left[\begin{array}{ccc}
230 & -7 & -7 \\
-7 & 230 & -7 \\
-7 & -7 & 230
\end{array}\right]\,
\begin{pmatrix}
n_{\Delta_1} \\ 
n_{\Delta_2} \\ 
n_{\Delta_3}
\end{pmatrix}\,\frac{6}{T^2}
\end{equation}
\\
\noindent
Using Eq.~\eqref{eq:SM_asymmetric_densities}, we can express the chemical potentials $\mu_{e_{\alpha}}$ in terms of the flavor asymmetries $n_{\Delta e_{\alpha}}$ and relate them to $n_{\Delta\alpha}$ as

\begin{equation}
\begin{pmatrix}
n_{\Delta e_1} \\ 
n_{\Delta e_2} \\ 
n_{\Delta e_3}
\end{pmatrix}
\,=\,
-\frac{1}{711}\,
\left[\begin{array}{ccc}
230 & -7 & -7 \\
-7 & 230 & -7 \\
-7 & -7 & 230
\end{array}\right]\,
\begin{pmatrix}
n_{\Delta_1} \\ 
n_{\Delta_2} \\ 
n_{\Delta_3}
\end{pmatrix}
\end{equation}
\\
\noindent
Hence, in this model and for $T_{sph}\ll T\ll 10^{8}\,\GeV$, the total baryon $n_B$ and the SM lepton $n_{L_{\text{SM}}}$ asymmetries are related to $n_{B-L_{\text{SM}}}$ by

\begin{align}\label{eq:baryon asymmetry FIBG}
n_B\,=\,\frac{22}{79}\,n_{B-L_{SM}}\,=\,\frac{22}{79}\,\sum_{\alpha=1}^3n_{\Delta\alpha}
\\
n_{L_{SM}}\,=\,-\frac{57}{79}\,n_{B-L_{SM}}\,=\,-\frac{57}{79}\,\sum_{\alpha=1}^3n_{\Delta\alpha}
\end{align}

\noindent
This result agrees with that obtained in reference \cite{Shuve_2020}.
\\
\end{appendix}

\bibliographystyle{JHEP}
\bibliography{bibliography}

\end{document}